\documentclass[%
 reprint,
 superscriptaddress,
 amsmath,amssymb,
 aps,
]{revtex4-2}

\usepackage{xcolor} 
\usepackage{braket}
\usepackage{ulem} 
 	\usepackage{graphicx}
	\usepackage[caption=false]{subfig}
	\usepackage[export]{adjustbox}

\usepackage{soul}

\usepackage{graphicx}
\usepackage{dcolumn}
\usepackage{bm}
\usepackage{textgreek}
\usepackage[version=4]{mhchem}

\begin{document}
\preprint{APS/123-QED}
\title{Exciton optics, dynamics and transport in atomically thin semiconductors}
\author{Raul Perea-Causin}
\thanks{These two authors contributed equally.}
\author{Daniel Erkensten}
\thanks{These two authors contributed equally.}
\affiliation{Department of Physics, Chalmers University of Technology, 412 96 Gothenburg, Sweden}
\author{Jamie M. Fitzgerald}
\author{Joshua J. P. Thompson}
\author{Roberto Rosati}
\author{Samuel Brem}
\affiliation{Department of Physics,
Philipps-Universität Marburg, 35032 Marburg, Germany}
\author{Ermin Malic}
\affiliation{Department of Physics,
Philipps-Universität Marburg, 35032 Marburg, Germany}
\affiliation{Department of Physics, Chalmers University of Technology, 412 96 Gothenburg, Sweden}

\begin{abstract}
Atomically thin semiconductors such as transition metal dichalcogenide (TMD) monolayers exhibit a very strong Coulomb interaction, giving rise to a rich exciton landscape. This makes these materials highly attractive for efficient and tunable optoelectronic devices.
In this article, we review the recent progress in the understanding of exciton optics, dynamics and transport, which crucially govern the operation of TMD-based devices.
We highlight the impact of hBN-encapsulation, which reveals a plethora of many-particle states in optical spectra, and we outline the most novel breakthroughs in the field of exciton-polaritonics.
Moreover, we underline the direct observation of exciton formation and thermalization in TMD monolayers and heterostructures in recent time-resolved ARPES studies. We also show the impact of exciton density, strain and dielectric environment on exciton diffusion and funneling.
Finally, we put forward relevant research directions in the field of atomically thin semiconductors for the near future.
\vspace{1cm}
\end{abstract}

\maketitle


Atomically thin semiconductors have emerged in the last decade as a platform for investigating quantum many-body phenomena and as promising candidates for novel optoelectronic applications\,\cite{Wang18, Mueller18, huang2022excitons, ciarrocchi2022excitonic, regan2022emerging, Novoselov16}.
In particular, monolayers of semiconducting transition metal dichalcogenides (TMDs), including the extensively studied MoS\textsubscript{2}, MoSe\textsubscript{2}, WS\textsubscript{2}, and WSe\textsubscript{2}, display weak dielectric screening due to their truly two-dimensional character. Concretely, the electric field lines between two charge carriers confined in the TMD monolayer extend largely outside the material into the surrounding dielectrics, which generally possess a small permittivity.
The resulting strong Coulomb interaction gives rise to rich exciton physics in TMD monolayers.
Excitons, i.e. tightly bound electron-hole pairs, dominate the optical response in these materials and show large binding energies and oscillator strengths\,\cite{Chernikov14}. 

The multi-valley band structure and sizable spin-orbit coupling in TMDs result in an abundance of optically active (bright) and inactive (dark) exciton states with different spin-valley configurations\,\cite{Yu15, Malic18, Wang18}.
Taking advantage of their two-dimensional character, TMD monolayers can be vertically stacked to form van der Waals (vdW) heterostructures, which allows for the creation of complex materials with designed functionalities and opens up a new venue for exploring intriguing many-body physics. In particular, the exciton landscape is extended to interlayer excitons, where the constituent electrons and holes are spatially separated, i.e. they are localized in different TMD layers\,\cite{Novoselov16, Rivera15, huang2022excitons, regan2022emerging}.
While excitons dominate the optical response in TMD monolayers and heterostructures at low or moderate carrier densities, higher-order many-particle complexes including trions, biexcitons, and polaritons become crucial in different technologically relevant regimes, i.e. in the presence of doping, at high excitation densities, and in optical cavities, respectively.

In order to design efficient and tunable optoelectronic devices based on TMDs, the fundamental processes governing the device operation must be well understood. These processes can be grouped into three categories---optics, dynamics and transport---which include (i) optical absorption and emission processes, (ii) exciton formation, thermalization and recombination dynamics, and (iii) exciton diffusion and funneling (cf. Fig.\,1).
In this article, we review the recent developments in the understanding of exciton optics, dynamics, and transport in semiconducting TMD monolayers and heterostructures. In particular, we focus on the four most studied TMDs, i.e. MoS\textsubscript{2}, MoSe\textsubscript{2}, WS\textsubscript{2}, and WSe\textsubscript{2}.
In Section I, we describe optical signatures of excitons and highlight the crucial impact of hBN-encapsulation, which reveals the fine structure of optical spectra with a multitude of many-particle states (cf. Fig.\,1A).
Furthermore, we review the recent advances in the extensive field of exciton-polaritons in monolayer TMDs and heterostructures.
In Section II, we discuss the progress in exciton dynamics including exciton formation, thermalization, and recombination (cf. Fig.\,1B). Here, the  focus lies on the direct observation of momentum-dark excitons in time-resolved ARPES measurements as well as interlayer exciton formation and recombination of moir\'{e} excitons. In Section III, we summarize the recent developments in exciton transport, emphasizing the impact of exciton density, strain and dielectric environment on exciton diffusion and funneling (cf. Fig.\,1C). Finally, we highlight the most promising avenues and pressing challenges for prospective research in this field.


\begin{figure*}[t!]
    \centering
  \includegraphics[width = \linewidth]{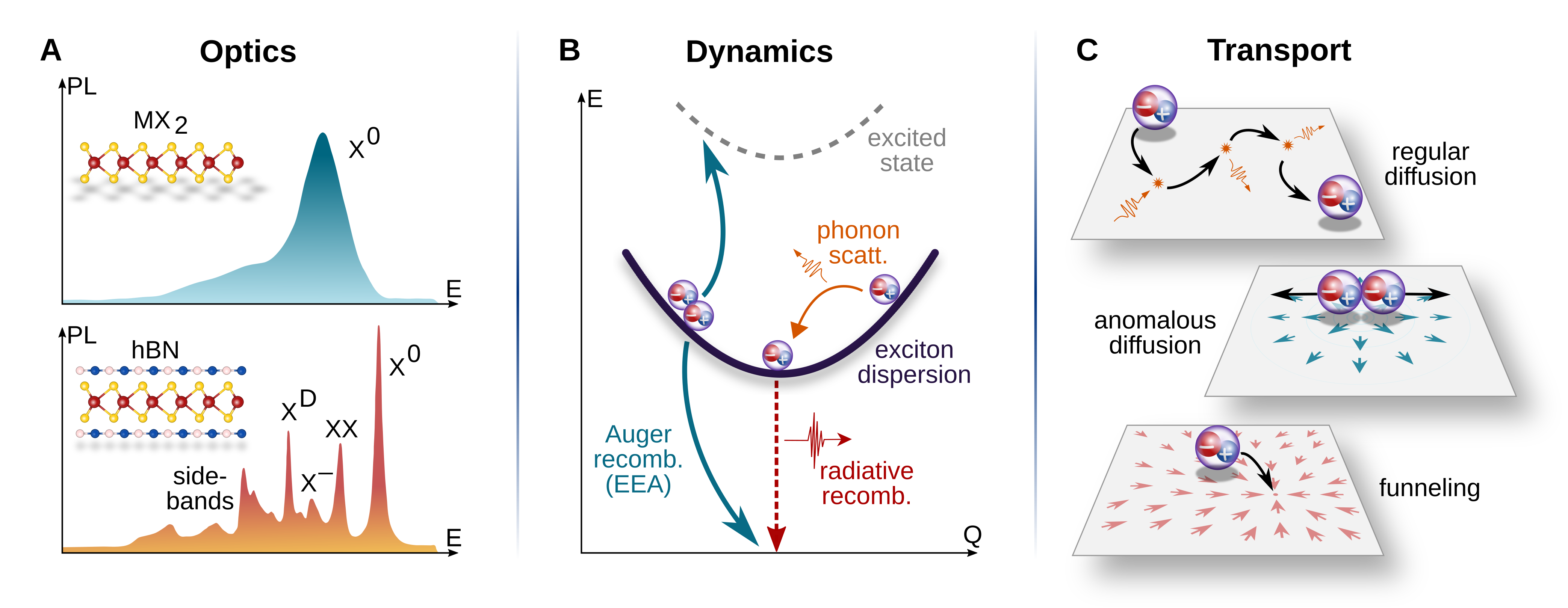}
    \caption{Schematic illustration of the three focus areas of the review. 
    \textbf{A}: Exciton optics with a multitude of resonances in photoluminescence spectra of hBN-encapsulated TMDs, including signatures of neutral (X$^0$) and dark (X$^\text{D}$) excitons, as well as trions (X$^-$) and biexcitons (XX).
    \textbf{B}: Exciton dynamics illustrating thermalization via exciton-phonon scattering and recombination via Auger scattering (exciton-exciton annihilation) and radiative channels.
    \textbf{C}: Exciton transport including regular diffusion---dominated by exciton-phonon scattering---, anomalous diffusion---governed by exciton-exciton repulsion---, and exciton funneling, where the arrows in the plane illustrate the direction of the drift current.
   }
    \label{fig:1}
\end{figure*}

\section{Exciton Optics}
The close relationship between optics and excitonics represents the primary and most versatile avenue for probing and manipulating excitons in transition metal dichalcogenides\,\cite{Wang18, Mueller18}. Excitons in semiconducting TMDs are typically generated via optical excitation\,\cite{mak10, Splendiani10}, where an incident photon leads to the generation of a Coulomb-bound electron-hole pair forming an exciton. In this section we discuss recent advances in exciton optics, which have been enabled by hBN-encapsulation of TMD samples. The sharp spectral lines resulting from the hBN-encapsulation allowed to reveal a rich excitonic substructure of bright and dark excitons, as well as higher-order charge complexes such as trions and biexcitons. Besides the exciton physics in TMD monolayers, we also outline how optics can be used to probe the nature of excitons in TMD-based heterostructures.
Furthermore, we explore how the exciton-light coupling can be enhanced by structuring the surrounding dielectric medium, e.g. in cavities, leading to the formation of exciton-polaritons.
 
\subsection{Bright and dark exciton signatures}
As the research field of TMDs has grown, fabrication and device design has steadily improved, with modern device architectures being of exceptionally high quality. This has been, in part, achieved through the encapsulation of TMDs with hexagonal boron nitride (hBN), an inert, two-dimensional wide bandgap insulator. The hBN efficiently protects the TMD from typically used substrates, which generally possess many defects, dangling bonds and a rough surface that creates dielectric inhomogeneities\,\cite{Uchiyama19, Raja19, dean2010boron}. 
Therefore, hBN-encapsulated samples are better protected from disorder effects which hamper the generation and detection of charge complexes\,\cite{Ye18}. This has triggered the optical detection of various excitonic and higher-order charge complex features in recent studies. In the following we will discuss signatures of bright and dark excitons which have been revealed through the improved quality of sample structures. 

Absorption measurements on TMDs can directly probe the energy of optically active excitons\,\cite{mak2012control}. In addition to the spectrally lowest 1s exciton, higher-order s-states\,\cite{goryca2019revealing}, as well as signatures from the optically dark p-states\,\cite{berghauser2016optical}, can be observed. Moreover, the spectral broadening of absorption resonances directly reflects both radiative and non-radiative dephasing processes\,\cite{Selig16, gupta2019fundamental, brem19b, fang2022room, Katsch20, erkenstenxx}.
Another important control parameter in absorption experiments is the optical polarization of the incident light. In TMDs, optical excitation by left- and right-handed circularly polarized light generates excitons at the K and K' valleys in the electronic dispersion, respectively\,\cite{yao2008valley, Xiao12, mak2012control, zeng2012valley}. This has important implications for both valleytronics\,\cite{lee2016electrical, schaibley2016valleytronics} and, due to the spin-valley locking in TMDs\,\cite{lu2019optical}, spintronics applications\,\cite{gong2013magnetoelectric, rostami2015valley, eginligil2015dichroic}.

\begin{figure*}[t!]
    \centering
  \includegraphics[width = \linewidth]{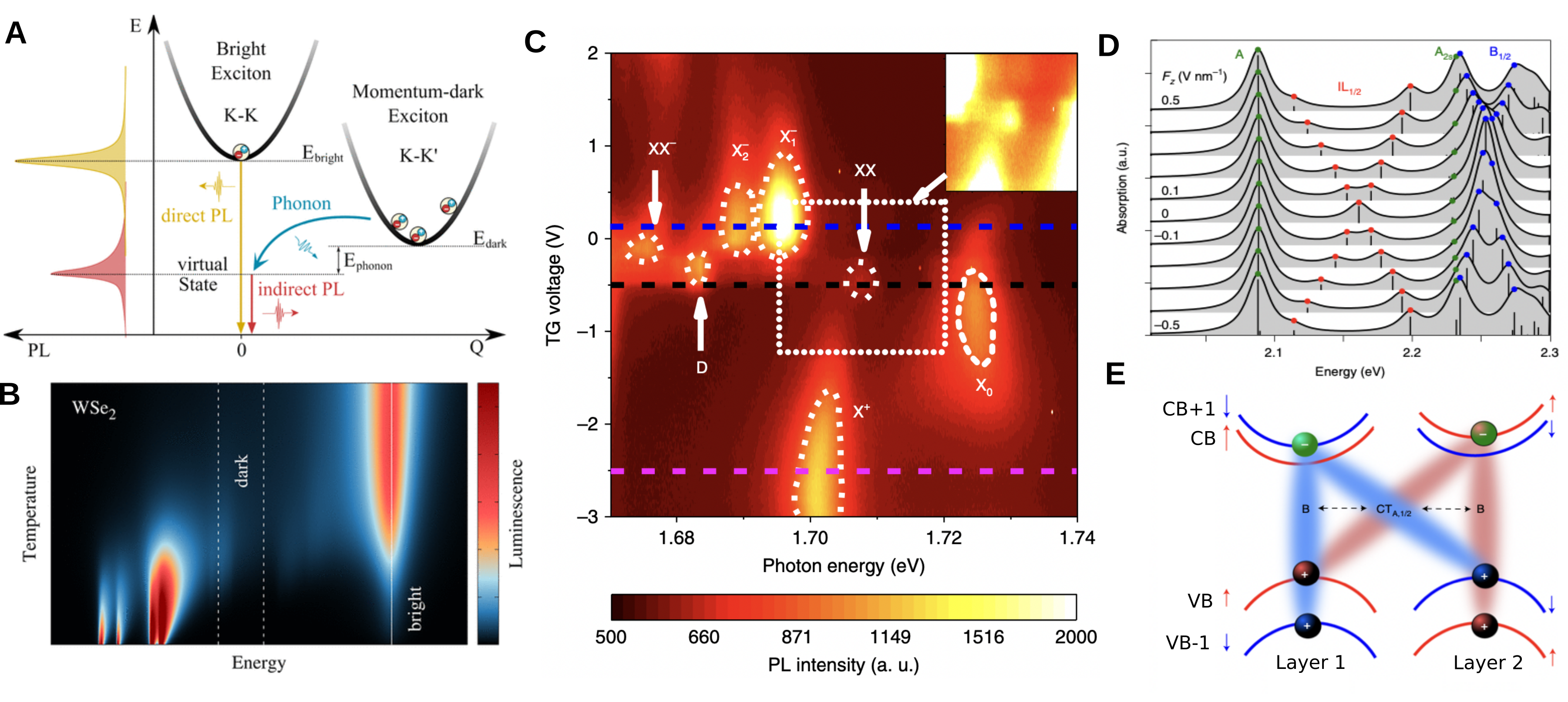}
    \caption{
    Optical signatures of excitons and higher-order many-body complexes in TMD monolayers and heterostructures.
    \textbf{A}: Sketch of the excitonic center-of-mass dispersion for a tungsten-based TMD (with energetically lower dark excitons), the direct and indirect recombination channels, and the expected PL signal.
    \textbf{B}: Temperature-dependent PL spectrum calculated for a \ce{WSe2} monolayer demonstrating the appearance of phonon sidebands at low energies.
    \textbf{C}: 
    PL spectrum for a hBN-encapsulated \ce{WSe} layer as a function of applied gate voltage, revealing a plethora of charged many-particle complexes.
    \textbf{D}: Calculated absorption spectrum (grey, shaded) and experimental peak positions (coloured dots) demonstrating electric field tuning of the hybrid interlayer exciton resonance in bilayer \ce{MoS2}.
    \textbf{E}: Schematic of the band alignment of bilayer \ce{MoS2} at finite electric field.
    Panels adapted with permission from: \textbf{A}-\textbf{B}, Ref.\,\cite{Brem20}, ACS under a Creative Commons License; \textbf{C}, Ref.\,\cite{li2018revealing}, Springer Nature Ltd. under a Creative Commons license; \textbf{D}-\textbf{E}, Ref.\,\cite{peimyoo2021electrical}, Springer Nature Ltd.}
    \label{fig:2}
\end{figure*}

Excitons posses a finite lifetime\,\cite{palummo15} before undergoing radiative decay and emitting a photon. Between their creation and annihilation, however, phonon- or Coulomb-mediated scattering processes allow the generated excitons to relax into some new distribution\,\cite{Brem18, trovatello20}. By measuring the optical response from the radiative recombination of these excitons in photoluminescence (PL) experiments, it is possible to uncover additional information on excitonic processes in the TMD.
In particular, PL spectra can exhibit unique signatures from the indirect recombination of dark excitons\,\cite{lindlau2018role, Brem20, Zhang15}.
Tungsten-based TMDs (such as WS$_2$ and WSe$_2$) possess energetically lower-lying momentum-dark exciton states (cf. Fig.\,\ref{fig:2}A), comprised of an electron and hole in different valleys in the Brillouin zone\,\cite{Malic18, Deilmann19, Selig16}. The large center-of-mass momentum of these excitons forbids direct radiative recombination. However, by  simultaneous interaction with a phonon, momentum conservation can be fulfilled and the dark exciton can indirectly emit a photon, leaving behind a clear signal in the PL spectra. This process is known as phonon-assisted photoluminescence\,\cite{Li19, Brem20,Lindlau17, lindlau2018role, funk2021spectral} and results in the formation of phonon sidebands, cf. Fig.\,\ref{fig:2}A. The probability of this higher-order process is significantly lower than the direct emission from a bright exciton. However, at sufficiently low temperatures, the energetically lowest dark excitons carry almost the entire population, which compensates for the much lower emission probability and results in a strong signal in the PL. Thus, in molybdenum-based TMDs, whose momentum-dark states are located higher in energy, no phonon sidebands are observed. The temperature-dependent PL spectrum for monolayer WSe$_2$\,\cite{Brem20} is shown in Fig.\,\ref{fig:2}B, where at high temperatures, only emission from the bright exciton can be seen. In contrast, at temperatures below 50 K, the indirect emission from dark excitons dominates the PL. Since here the emission of a phonon is involved, the phonon sidebands are red-shifted in energy with respect to the position of dark excitons (dashed white lines). Multiple resonances are observed due to different optical and acoustic phonons involved in this process. These phonon sidebands have been observed both for dark excitons\,\cite{Brem20,he20, Liu20,funk2021spectral}, where they dominate the PL spectra, and bright excitons, where they are hidden in an asymmetric broadening of the spectral linewidth\,\cite{Brem20,Christiansen17}. 

Note that besides momentum-dark excitons discussed so far, there exist also spin-dark excitons consisting of an electron and hole with an opposite spin. These excitons can be brightened under the application of an external magnetic field\,\cite{robert17,peng19, feierabend20} or by changing the angle of the detector/polarization\,\cite{wang17}. In the latter case, the magnetic component of the electromagnetic field points out-of-plane and hence couples the opposite spins allowing for the two charge carriers to recombine.

Excitons are characterized by a band structure with an effective mass determined by the excitonic center-of-mass motion (cf. Fig.\, \ref{fig:2}A). PL and absorption spectroscopy can only probe the small-momentum range within the light cone directly and hence do not provide any momentum-resolved information. Instead, recent studies\,\cite{Madeo20, lee2021time, Wallauer21, schmitt21, Dong21} have employed angle-resolved photoemission spectroscopy (ARPES) measurements, which can be used to probe the excitonic band structure more directly. In this technique, an initial pump pulse generates excitons before a high-energy photon ejects electrons from the conduction band via the photoelectric effect.
While ARPES has the advantage of resolving the electron population in the whole Brillouin zone, it is not directly clear whether the ejected conduction-band electron was bound to a hole (forming an exciton). Nevertheless, the excitonic nature of the measured signal can be determined with complementary PL or absorption spectra. Concretely, the agreement between the bandgap extracted from ARPES and the exciton resonance energy in optical spectra provides direct evidence of the excitonic nature of the ARPES signal.
In Section II we explore how time-resolved ARPES measurements can be used to probe the exciton dynamics and in particular the thermalization between bright and dark excitons\,\cite{Wallauer21, Madeo20}.

\subsection{Trion and biexcion signatures}
The optical response of TMD monolayers is generally dominated by excitons at low and moderate excitation density in weakly doped samples. At elevated excitation densities and in the presence of doping, higher-order charge complexes become relevant.
In general, the optical signatures of different many-particle charge complexes are only separated by up to  tens of meV. For TMDs deposited on typical substrates such as SiO\textsubscript{2}, the inhomogeneous broadening of spectral resonances on the order of several tens of meV makes it challenging---if not impossible---to optically resolve these signatures. Remarkably, encapsulation of the TMD monolayer by hBN leads to a narrowing of the inhomogeneous linewidth, unveiling a plethora of signatures in absorption and PL spectra arising from complex recombination processes of many-body compounds.
In this section we report on recent experiments probing trion and biexciton features in TMDs.

Trions, formed from an exciton and an additional electron or hole, have been observed in absorption and PL measurements\,\cite{mak13, ross13, Plechinger16, courtade17, sidler2017fermi}. Biexcitons\,\cite{hao2017neutral, steinhoff2018biexciton, katsch2020optical}---consisting of two bound electron-hole pairs--- and charged biexcitons\,\cite{barbone18, Ye18, paur2019electroluminescence, li2018revealing}---biexciton plus an electron/hole--- exhibit also features that could be observed in optical spectroscopy. In order to obtain a significant trion signal, the TMD must be doped. This can be achieved via intrinsic doping\,\cite{zhumagulov2020trion, rivera2021intrinsic} or by the application of an external electric field\,\cite{li2018revealing}. Owing to the Coulomb interaction between the exciton and the electron/hole, the trion has a binding energy of $\sim 20-30$ meV and is thus red-shifted from the corresponding exciton energy\,\cite{mak13,ross13}.
Moreover, absorption measurements show that the energy splitting between trion and exciton resonances increases with doping, providing a signature of the polaronic character of the exciton in the sea of doping charges\,\cite{sidler2017fermi,suris2001excitons,efimkin2017many}. For this reason, trion and exciton resonances are commonly denoted attractive and repulsive polaron branches in absorption measurements.
Recent studies\,\cite{liu19a, he20} reported also the existence of dark trions at low temperatures in WSe$_2$. Similar to the excitonic case, they become visible via interaction with phonons. In these studies, the doping was controlled by an applied electric field resulting in three distinct regions of the PL dominated by dark excitons (low/no doping), dark positively-charged trions (p-type doping) and dark negatively-charged trions (n-type doping). 

Figure\,\ref{fig:2}C shows the PL spectrum around the bright exciton energy as a function of applied gate voltage for a hBN-encapsulated WSe$_2$ monolayer\,\cite{li2018revealing}. Due to the sharp lines, one can find a multitude of signatures including the neutral exciton (labelled X$_0$), positive trion (X$^+$), negative intervalley (X$_1^-$) and intravalley (X$_2^-$) trions, biexciton (XX), and negatively-charged biexciton (XX$^-$). There is also clear emission from bound defect excitons (D). These states have been shown to enhance the PL intensity of the free exciton\,\cite{tongay2013defects} and they are important for the realization of single-photon emitters\,\cite{Tonndorf15, He15,  Thompson20, parto2021defect}. Through hBN-encapsulation recent studies could even resolve the excited trionic 2s states\,\cite{arora2019excited, wagner20, liu2021exciton}.

\subsection{Exciton optics in TMD heterostructures}
The two-dimensional nature of TMD monolayers encourages the vertical stacking of multiple TMD layers. The resulting structure, held together through the van der Waals (vdW) interaction, inherits properties of the constituent monolayers. TMD homobilayers have typically an indirect bandgap due to a large hybridization of the states at the $\Lambda$ and $\Gamma$ valleys\,\cite{ruiz2018hybrid}. Unlike in monolayers, whose PL is dominated by the energetically lowest direct optical transition at room temperature, homobilayers\,\cite{chu2015electrically} possess a rich structure of momentum indirect and direct excitons\,\cite{lindlau2018role}, whose optical properties can be tuned via strain\,\cite{Conley13, leisgang2020giant} or by an electric field\,\cite{leisgang2020giant}.
Heterobilayers, formed from stacking two different TMD monolayers, can play host to interlayer excitons\,\cite{Rivera15}, where the constituent electron and hole reside in different layers. The spatial separation of the electron and the hole strongly quenches the radiative recombination of these excitons and as such they possess long lifetimes\,\cite{Miller17, jiang18}, which is important for transport phenomena\,\cite{huang2019robust} and energy transfer processes in e.g. solar cells\,\cite{furchi2018device}.

Interlayer excitons carry a permanent out-of-plane dipole moment, allowing their energy to be tuned via an electric field\,\cite{jauregui19, kiemle2020control, Kis19}. In PL and absorption measurements this leads to a shift in the exciton resonance energy. Interestingly, interlayer excitons are also predicted to exist in homobilayers\,\cite{deilmann2018interlayer}.
Moreover, these interlayer excitons strongly hybridize with the intralayer state, inheriting some of their oscillator strength, and are thus visible even at room temperature\,\cite{deilmann2018interlayer, peimyoo2021electrical}.
It was recently shown that by varying the gate voltage across the structure, the interlayer exciton resonance can be significantly tuned\,\cite{peimyoo2021electrical} (cf. Fig.\,\ref{fig:2}D). Concretely, the degeneracy of the two interlayer excitons is lifted and their resonances move closer to the A1s and A2s intralayer excitons, respectively (cf. Fig.\,\ref{fig:2}E). 

\begin{figure*}[t!]
    \centering
  \includegraphics[width = \linewidth]{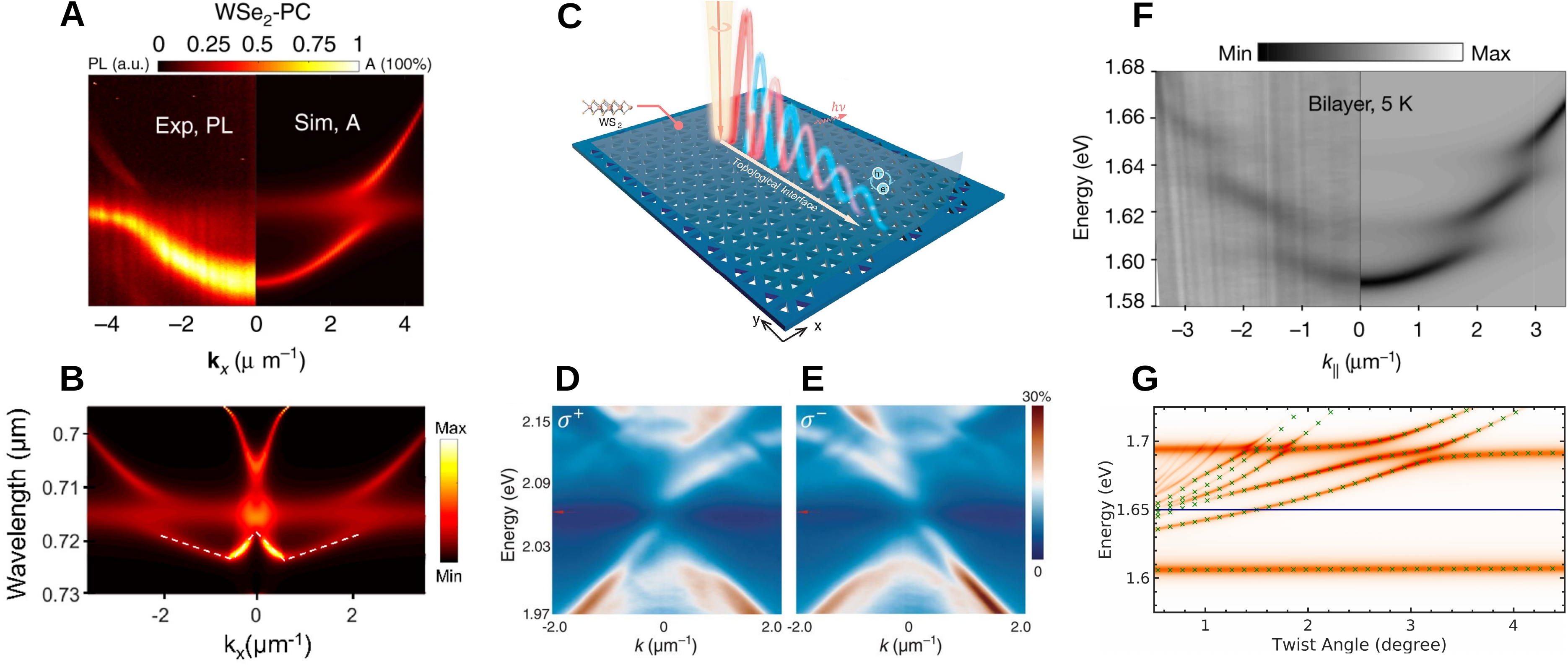}
    \caption{Exciton-polaritons in TMD monolayers and heterostructures.
    \textbf{A}: Experimental and simulated PL spectrum for a coupled monolayer TMD and photonic crystal slab in the strong coupling regime.
    \textbf{B}: Simulated "W-shaped" dispersion of exciton-polaritons supported by a monolayer TMD on a metasurface, which is tunable via lattice periodicity and slab thickness.
    \textbf{C}: Illustration of helical topological interface TMD exciton-polaritons.
    \textbf{D/E}: Angle-resolved reflectance spectra of the interface polaritons excited by right/left-hand circularly polarized light.
    \textbf{F}: First experimental evidence of strong coupling between two hybrid interlayer moir\'{e} excitons and a microcavity optical mode for a \ce{MoSe2}/\ce{WS2} heterostructure, measured in angle-resolved reflection spectra.
    \textbf{G}: Twist-angle dependence of polariton dispersion and absorption for a \ce{MoSe2}/\ce{WSe2} heterostructure in a microcavity.
    Panels adapted with permission from: \textbf{A}, Ref.\,\cite{zhang2018photonic}, Springer Nature Ltd. under a Creative Commons license; \textbf{B}, Ref.\,\cite{chen2020metasurface}, ACS; \textbf{C}-\textbf{E}, Ref.\,\cite{liu2020generation}, AAAS; \textbf{F}, Ref.\,\cite{zhang2021van}, Springer Nature Ltd.; \textbf{G}, Ref.\,\cite{Fitzgerald}, ACS under a Creative Commons license.
    }
    \label{fig:3}
\end{figure*}

In recent years, twisted TMD heterostructures have ignited substantial interest and given rise to the rapidly evolving field of semiconducting moir\'{e} superlattices\,\cite{huang2022excitons}. Here, by tuning the twist angle between two or more stacked TMD layers, a new long-range periodicity can be created leading to novel electronic and excitonic properties\,\cite{yu2017moire, rivera2018interlayer, Seyler19, Tran19}. 
The long-range moir\'{e} pattern manifests as a periodic network of potential wells\,\cite{wu2018theory, ruiz2019interlayer, Brem20c, hagel2021exciton}, trapping excitons when sufficiently deep. This trapping leads to the formation of distinct moir\'{e} exciton subbands, tens of meV below the free exciton energy level. Optically, this can be observed in the twist-angle dependent fine structure of both inter-\,\cite{Tran19} and intralayer\,\cite{zhang2018moire, Brem20c} excitons.
In addition, the structure of moir\'{e} excitons can be analyzed in detail in ARPES measurements. Concretely, the spatial extension of the exciton can be extracted from the momentum distribution of the ARPES signal.
In particular, interlayer excitons in WSe\textsubscript{2}/MoS\textsubscript{2} structures with a twist angle of 2{\textdegree} were recently found to be localized within the 6 nm moir\'{e} unit cell\,\cite{karni2022structure}. Moreover, in the same structure but with a 10{\textdegree} twist angle, analogous experiments report the delocalized nature of interlayer excitons and their unique momentum fingerprint, which directly reflects the mini Brillouin zone of the moir\'{e} superlattice\,\cite{schmitt21}.

Recent advances have shown that TMD-based moir\'{e} heterostructures undergo significant lattice reconstruction at low angles, leading to an even more complex network of trapped exciton domains\,\cite{rosenberger2020twist, andersen2021excitons} and giving rise to ferroelectric\,\cite{enaldiev2021piezoelectric} and strain-induced effects\,\cite{edelberg2020tunable}. Analogously to monolayer systems, homo- and heterobilayers exhibit higher-order charge complexes, such as moir\'{e} trions\,\cite{brotons2021moire, wang2021moire, liu2021excitonic, liu2021signatures}, which appear in optical spectra as an additional fine structure of the trion peak. For more details on this exciting field of exciton research, we refer to a very recent review on moir\'{e} exciton effects\,\cite{huang2022excitons}.

\subsection{Exciton-polaritons}
The interaction of excitons with light can be qualitatively modified by structuring the surrounding dielectric environment\,\cite{kavokin2017microcavities}. TMD monolayers and heterostructures are particularly relevant in the domain of nanophotonics as they are chemically stable at ambient conditions, do not suffer from lattice mismatch, and their exciton energy lies in the technologically relevant spectral region of visible and near-infrared, allowing for incorporation with a wealth of nanophotonic devices\,\cite{hu2020recent}. TMD nanophotonics can be divided into the weak and the strong coupling regime depending on the strength of the exciton-light interaction in comparison to the optical and material-based decay channels. In the former, losses dominate and the (dressed) exciton remains a well-defined eigenmode of the system with a modified lifetime (Purcell effect). In other words, spectra can remain qualitatively unchanged, but exciton emission and absorption are enhanced/diminished via control of the local photonic density of states. For example, the low quantum yield of TMDs can be compensated by coupling to a suitably designed nanostructure to boost Raman and PL signals\,\,\cite{galfsky2016broadband,fang19}. This not only can aid optical characterization of 2D materials\,\,\cite{lee2017single}, but also has applications for ultra-compact optoelectronic devices based on excitons\,\cite{krasnok2018nanophotonics,cotrufo2019enhancing}. In a further example, it has been demonstrated that the weak absorption of monolayers ($\lesssim 10\%$)\,\,\cite{bernardi2013extraordinary} can be greatly increased to near $100\%$ using the Salisbury screen setup\,\,\cite{epstein2020near,horng2020perfect}, which could have utility for ultra-thin photovoltaics. In this section, we concentrate on TMDs in the strong-coupling regime\,\cite{schneider2018two,krasnok2018nanophotonics,hu2020recent,zhao2021strong,al2022recent}, focusing particularly on utilizing nanophotonics for sophisticated control of TMD exciton-polaritons, and the recent development of polariton twistronics. 

The light-matter coupling strength of TMD heterostructures integrated into high-quality and small-volume optical microcavities can exceed both material dissipation and radiative decay from the cavity (both typically on the order of few to tens of meV), leading to coherent energy transfer between the exciton and cavity mode\,\cite{kavokin2017microcavities}. In this case, the system is in the strong-coupling regime and is characterized by the formation of hybrid quasi-particles, known as exciton-polaritons. These new eigenmodes of the combined light-exciton system inherit properties from their constituent parts, potentially combining the spatial coherence, small effective mass, and long propagation lengths of photons with the tunability and nonlinearity of material-based excitations\,\cite{sanvitto2016road}. In the frequency domain, the strong-coupling regime manifests as a splitting of the polariton energies known as Rabi splitting (cf. Fig.\,\ref{fig:3}A), which is determined by the strength of the light-exciton coupling and can be directly observed in angle-resolved linear optical spectroscopy\,\cite{kavokin2017microcavities}. The large oscillator strength of monolayer TMD excitons leads to an impressive Rabi splitting (up to $\sim 50$ meV in a typical dielectric microcavity\,\cite{liu2015strong}) in comparison to traditional polariton platforms such as \ce{GaAs} quantum well microcavities ($\sim 4$ meV for a single quantum well in a microcavity\,\cite{brodbeck2017experimental}). Furthermore, the large binding energy allows for robust room-temperature polaritonics. The study of exciton-polaritons in TMDs has led to a wealth of interesting phenomena, such as condensation of exciton-polaritons at cryogenic temperatures\,\cite{anton2021bosonic}, polariton lasing\,\cite{zhao2021ultralow}, trapped\,\cite{shan2021spatial} and Bloch polaritons\,\cite{lackner2021tunable}, motional narrowing\,\cite{wurdack21} and nonlinear polariton parametric emission\,\cite{zhao2022nonlinear}.

The first demonstrations of the strong coupling regime in TMD monolayers used conventional dielectric Fabry-Perot cavities built from high-reflectivity distributed Bragg reflectors (DBR)\,\cite{liu2015strong,dufferwiel2015exciton}, hybrid DBR plus silver mirror\,\cite{flatten2016room}, and all-metal Fabry-Perot cavity built from silver mirrors\,\cite{wang2016coherent}. Development of state-of-the-art fabrication methods allows for routine and precise construction on the nanoscale, meaning that the near- and far-field properties of light can be manipulated to form sophisticated cavities with controllable properties such as quality factor, field enhancement, polarization and topology. For instance, Zhang et al. reported on a monolayer TMD placed on a $100$ nm thick 1D photonic crystal slab\,\cite{zhang2018photonic}. Recording both angle-resolved reflectance and PL, convincing evidence of strong coupling physics has been found up to room temperature between \ce{WS2} excitons and a guided mode supported by the structured slab (cf. Fig.\,\ref{fig:3}A). In another notable work\,\cite{chen2020metasurface}, exciton-polaritons were studied in a \ce{WSe2} monolayer on a 2D \ce{SiN} metasurface with a square lattice of holes. A strong directional polariton emission was measured in the far field due to the sub-wavelength structuring. Furthermore, a computational study demonstrated a highly tunable Rabi splitting, directional emission and polariton dispersion via meta-atom engineering, such as changing lattice periodicity and slab thickness. For instance, excitons can couple to a photonic crystal mode with a "W-shaped" dispersion (cf. Fig.\,\ref{fig:3}B). In this case, the polaritons will inherit a negative effective mass and group velocity from the optical mode at normal incidence. This can lead to exotic polariton transport\,\cite{arnardottir2017hyperbolic}. In another work, the strong coupling between excitons and a bound state in the continuum supported by a 1D photonic crystal slab was demonstrated\,\cite{kravtsov2020nonlinear}.

There is growing interest in exploring topological properties of polaritons\,\cite{karzig2015topological,liu2020generation,li2021experimental}. One strategy towards this goal is coupling excitons to topologically non-trivial photonic systems\,\cite{ozawa2019topological}, which could lead to tunable optical devices that are robust against disorder and defects. Helical topological polaritons have been demonstrated using a \ce{WS2} monolayer placed on a hexagonal photonic crystal\,\cite{liu2020generation} (cf. Fig.\,\ref{fig:3}C). While maintaining sixfold rotational symmetry, the unit cell can be expanded or shrunk to change the intercell and intracell nearest-neighbour couplings. This allows two bulk lattice regions to be engineered with different topological phases, with the boundary between the two supporting a topological interface mode\,\cite{ozawa2019topological}. This interface was found to support a topological interface polariton due to the strong coupling between the photonic interface mode and the \ce{WS2} exciton, or in an equally valid picture, as a consequence of the different topological phases of the bulk polariton bands in the two lattice regions. Crucially, the interface polaritons posses a helical nature, i.e. the group velocity depends on the pseudospin of the polariton, which can be directly accessed via the circular polarization of incident light (cf. Figs.\,\ref{fig:3}D and E). Ref.\,\cite{liu2020generation} demonstrates this unidirectional polariton transport in momentum space with reflection and PL measurements of the interface at temperatures of $160$-$200$ K. 

Several unique aspects of TMD exciton-polaritons have been uncovered in recent years. Because of the large exciton binding energy, the temperature dependence of the exciton energy (via the bandgap\,\cite{Odonnell91}) can be used as a material-based detuning over a wide range of temperatures. In an early demonstration for monolayer \ce{WS2}, the light-exciton composition of polaritons was modified over the range 110-230 K, and could be inferred from angle-resolved PL\,\cite{liu2017control}. As temperature was increased, the exciton energy was observed to red-shift, causing the lower polariton to shift from a photon-like state to an exciton-like state at small momenta. Strong coupling physics can also be modified through electric field induced gating of the TMD\,\cite{chakraborty2018control}. Here, by introducing electrostatically induced free carriers, the oscillator strength of the exciton decreases due to the screened Coulomb interaction between the constituent electrons and holes---this in turn leads to a reduced Rabi splitting. Furthermore, TMD exciton-polaritons were shown to inherit the valley physics of the constituent exciton\,\cite{chen2017valley,sun2017optical,dufferwiel2017valley}. Research on valley polaritonics has emerged as a popular subfield, and numerous combined nanophotonic and valleytronic architectures have been explored\,\cite{dufferwiel2018valley,qiu2019room,liu2021nonlinear}. In a notable example, it was shown that a condensate of exciton-polaritons in a hybrid monolayer \ce{MoSe2}-\ce{GaAs} microcavity preserves valley polarization of the pump laser better than in the linear regime due to the speed up of the relaxation dynamics from the excitonic reservoir\,\cite{waldherr2018observation}. The large exciton binding energy and oscillator strength of TMD monolayer excitons has also opened up opportunities to explore polaritonics beyond the 1s exciton, offering insights into many-body physics of TMDs\,\cite{sidler2017fermi} and the potential for highly nonlinear exciton-based optical devices\,\cite{barachati2018interacting,kravtsov2020nonlinear,tan2020interacting}. For example, the strong coupling between the 2s exciton in monolayer \ce{WSe2} and the optical modes of a Fabry-Perot cavity has been demonstrated\,\cite{gu2021enhanced}. Strong coupling with trions has also been reported\,\cite{dufferwiel2017valley,sidler2017fermi,lundt2017valley,lee2017electrical,cuadra2018observation}. Because of their charged and composite fermionic nature, trion polaritons possess strong nonlinear behaviour compared to neutral bosonic excitons\,\cite{emmanuele2020highly}. 

Recently, hybrid intra-interlayer moir\'{e} exciton-polaritons were explored in a H-stacked \ce{MoSe2}/\ce{WS2} heterostructure with a twist angle of $\sim 56.5^\circ$ placed in a $\lambda/2$ microcavity\,\cite{zhang2021van}. Here, a twist-dependent interlayer hybridisation means that the interlayer excitons inherit a large oscillator strength from the \ce{MoSe2}-based intralayer excitons at small twist angles\,\cite{Alexeev19}. It was found that the two bare moir\'{e} excitons coherently couple to the cavity mode to form three moir\'{e} polariton branches, which are measured with angle-resolved reflection and PL (cf. Fig.\,\ref{fig:3}F). In other words, the moir\'{e} polaritons in this system are a three-way coherent coupling between a cavity photon and the bare intra- and interlayer moir\'{e} excitons. Relative to monolayer exciton-polaritons, these hybrid moir\'{e} polaritons show negligible energy shift, smaller line-broadening, and much stronger reduction in the Rabi splitting with increasing excitation densities. These observations are attributed to the zero dimensionality of the moir\'{e} exciton and a consequent exciton blockade effect\,\cite{schneider2016exciton}. Localization of the exciton at the potential minima in each moir\'{e} supercell can be expected to lead to enhanced exchange and dipole-dipole interactions (via the interlayer component). This extra energy cost of adding an exciton to a moir\'{e} cell leads to a suppression of many-body effects and the exciton-photon coupling saturating at one exciton per moir\'{e} cell. This study represents an exciting first step towards ``polaritonic twistronics". Unfortunately, experiments are typically limited to a few fixed twist angles, while in theoretical studies one can map out the moir\'{e} exciton-polariton energy and absorption over a wide range of twist angles, as recently demonstrated for the specific case of the purely intralayer exciton-polaritons of a twisted type II \ce{MoSe2}/\ce{WSe2} heterostructure in a microcavity using a combined Wannier plus Hopfield method\,\cite{Fitzgerald} (cf. Fig.\,\ref{fig:3}G). Further theoretical studies have looked at possible topological transport properties of interlayer exciton-polaritons in a cavity\,\cite{yu2020electrically}, while a Bose-Hubbard model has been used to explore moir\'{e}-induced optical nonlinearities\,\cite{camacho2022moire}, and a quantum-electrodynamical extension of the Bethe-Salpeter method has been used to explore an interesting photon-induced reordering and mixing of intra- and interlayer excitons in an untwisted \ce{MoS2}/\ce{WS2} bilayer\,\cite{latini2019cavity}.


\section{Exciton dynamics}
\begin{figure*}[t!]
\includegraphics[width=\textwidth]{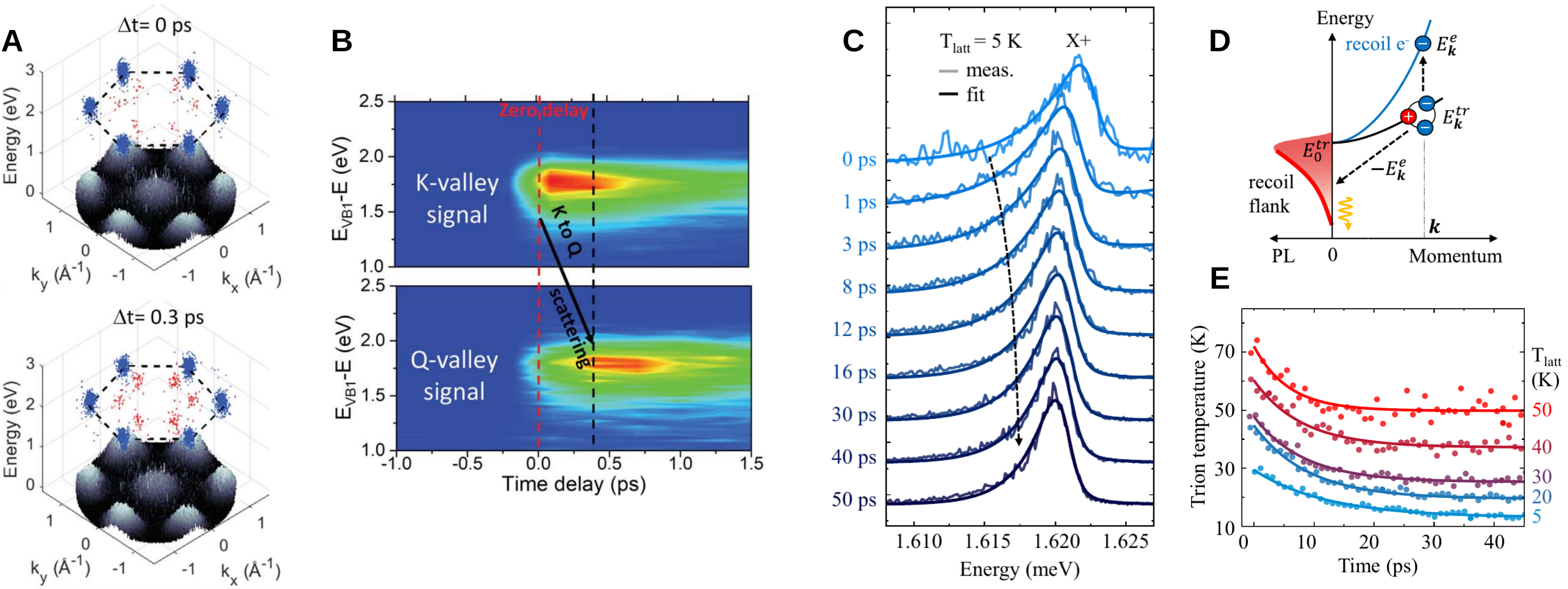}
\caption{Exciton and trion dynamics in monolayer TMDs.
\textbf{A}: ARPES maps of the electron occupation in the conduction and valence bands during (0 ps) and shortly after (0.3 ps) the optical excitation.
\textbf{B}: Time-resolved energy scan of the ARPES signal at the K and {\textLambda} (Q) valleys illustrating the thermalization of excitons into the dark K{\textLambda} states.
\textbf{C}: Time snapshots of the trion photoluminescence, displaying an asymmetric broadening that can be traced back to the trion temperature.
\textbf{D}: Schematic illustration of trion recombination and the electron recoil effect.
\textbf{E}: Time evolution of the extracted trion temperature for different lattice temperatures, demonstrating trion cooling.
Panels adapted with permission from: \textbf{A}-\textbf{B}, Ref.\,\cite{Madeo20}, AAAS; \textbf{C}-\textbf{E}, Ref.\,\cite{zipfel22}, APS.
} 
\label{fig:4}
\end{figure*}

The dynamics of exciton formation, thermalization, and recombination crucially determines the performance of TMD-based optoelectronic devices. For instance, the thermalization of excitons into long-lived dark states in tungsten-based TMDs results in the suppression of the PL, making these materials undesirable for light-emitting applications but suitable for light harvesting. Similarly, non-radiative recombination processes, such as exciton-exciton annihilation, can significantly limit the quantum yield. At moderate doping levels or large exciton densities, many-body complexes, such as trions or biexcitons, become relevant and dominate the thermalization and recombination dynamics.
Moreover, in TMD heterostructures, charge transfer arises as a major thermalization mechanism, and the exciton recombination dynamics becomes significantly influenced by the stacking and the twist angle of the constituent layers.
In this section, we summarize the recent progress in the understanding of these important processes.

\subsection{Exciton formation and thermalization}
Recent advances in two-photon photoemission spectroscopy (2PPS), also known as time-resolved ARPES (tr-ARPES), has enabled the direct observation of exciton formation and thermalization in energy, momentum and time\,\cite{Madeo20,Wallauer21,Dong21,kutnyakhov20,schmitt21}.
In this technique, an initial pump pulse photoexcites electrons from the valence band to the conduction band, resulting in the formation of excitons. After a certain time delay, the excitons' constituent electrons are photoemitted by a second pulse, and their energy and momentum are then measured by a hemispherical analyzer or a time-of-flight momentum microscope\,\cite{kutnyakhov20}.
The detected electrons directly reflect the exciton population and, importantly, their momentum configuration. Thus, ARPES allows for the direct visualization even of momentum-dark excitons, which are not directly accessible in PL or optical absorption spectra.

The formation of momentum-dark excitons in WSe\textsubscript{2} monolayers has been resolved in a very recent tr-ARPES study \cite{Madeo20}. First, electrons and holes at the K point are optically excited, leading to an almost immediate signal arising from photoemitted electrons with momentum K (cf. Fig.\,\ref{fig:4}A-B). After a delay of 0.4 ps, electrons with momentum {\textLambda} (also commonly denoted Q or {\textSigma} in literature) and an energy similar to those at the K point are detected. The measured energy of {\textLambda} electrons is significantly lower than the energy predicted by ab-initio calculations of the single-particle band structure.
However, due to their large binding energy, momentum-dark K{\textLambda} excitons (formed by a hole at the K point and an electron at the {\textLambda} point) have been predicted to reside energetically below the bright KK states\,\cite{christiansen19,Deilmann19,Selig18,Wallauer21,Malic18}.
In other words, the observed {\textLambda} electrons are bound to holes at the K valley and thus form momentum-dark K{\textLambda} excitons. The delay between the appearance of K and {\textLambda} electrons is therefore a direct reflection of the intervalley exciton-phonon scattering, which is responsible for the thermalization of the exciton population in the multi-valley bandstructure of TMDs\,\cite{Selig18,Wallauer21}. The measured delay of 0.4 ps thus quantifies the exciton-phonon scattering time at the considered temperature of 90 K.
Another experiment has been performed for monolayer WS\textsubscript{2}\,\cite{Wallauer21}, where the similar energy of KK and K{\textLambda} excitons was resolved and an intervalley scattering time of 16 fs at room temperature was extracted.
The definite proof that photoemitted electrons arise from bound exciton states relies on the direct visualization of the exciton wave function and the theoretical prediction of an inverted dispersion relation for the photoemitted electron \cite{christiansen19}. Both features have been successfully observed in monolayer\,\cite{man21} and bulk\,\cite{Dong21} WSe\textsubscript{2}. These recent studies investigating exciton formation with tr-ARPES confirm the importance of momentum-dark K{\textLambda} excitons in tungsten-based TMDs and provide a direct quantification of the intervalley exciton-phonon scattering time in these materials.

The exciton thermalization dynamics can also be probed indirectly via the optical absorption and emission characteristics of the material\,\cite{steinleitner17,berghauser18,rosati20b,ceballos16,Brem18,trovatello20}. 
Initial studies focused on resolving the valley and spin depolarization after a circularly-polarized optical excitation\,\cite{yu14,zhu14,selig19,xu21}.
Other studies have utilized indirect signatures of dark excitons in optical spectroscopy to track their formation.
For example, optical-pump THz-probe experiments can be performed to resolve the exciton dynamics by exploiting the temporal evolution of the 1s-2p exciton transition resonance\,\cite{pollmann15,steinleitner17}. In particular, the blue-shift of the 1s-2p THz resonance peak reflects the sub-picosecond relaxation of bright KK excitons into the lower-lying momentum-dark K{\textLambda} state, which displays a larger 1s-2p resonance energy\,\cite{berghauser18}.
The formation dynamics of dark excitons has also been indirectly resolved by tracking the emergence of phonon-assisted PL peaks at cryogenic temperatures\,\cite{rosati20b}.

The exciton formation can also be resolved by exploiting the population-induced changes in the optical absorption \,\cite{ceballos16,Brem18,trovatello20}. In particular, above-band optical excitation provides a unique fingerprint to track the relaxation cascade into the exciton ground state\,\cite{Brem18,trovatello20}.
Signatures of efficient phonon cascades that would enable this process have been recently observed\,\cite{Paradisanos21}.
The opposite process of exciton dissociation can occur via an external electric field\,\cite{massicotte18}
or by scattering with phonons\,\cite{perea21} and crucially limits the performance of TMD-based optoelectronic devices.

\subsection{Exciton recombination}
Having binding energies in the range of hundreds of meV, excitons govern the dynamics of electron-hole recombination in monolayer semiconductors and vdW heterostructures. Depending on the dielectric environment and excitation power, excitons mainly recombine radiatively by emitting photons or non-radiatively by other processes, e.g. via defect-assisted recombination or exciton-exciton annihilation \cite{Kumar14b, Zipfel20}.
Radiative recombination in monolayers has been extensively studied in the past\,\cite{palummo15,pollmann15,robert16,wang16}, with the hallmark being the characteristic temperature-dependence of the recombination time in tungsten- and molybdenum-based TMDs owing to the dominance of dark excitons in the first case and bright excitons in the second\,\cite{Zhang15,Selig18}.
More recently, the radiative lifetime has been found to be tunable with the thickness of the hBN-encapsulation layers due to the Purcell effect\,\cite{fang19}.

Electron-hole pairs can also recombine via non-radiative recombination processes, such as exciton-exciton annihilation (EEA) at elevated exciton densities \cite{sun14, Mouri14, yuan2015exciton, PRXGlazov, hoshi17, goodman2020substrate, Zipfel20, erkensten2021, Uddin22}, and defect-assisted recombination\,\cite{amani15,wang15,Zipfel20,lien19}. EEA crucially limits the efficiency of optoelectronic devices as it leads to a saturation of the quantum yield with increasing carrier density. Nevertheless, it also presents an opportunity for photon upconversion\,\cite{PRXGlazov,lin2021narrow}. In this Auger-like recombination process one exciton recombines by transferring its energy and momentum to another exciton.
In order for this mechanism to occur, energy- and momentum-conservation must be fulfilled. This implies that the efficiency of EEA is highly sensitive to the excitonic band structure, which contains the suitable final states. In fact, the EEA rate has been found to vary significantly with the dielectric environment \cite{hoshi17, Zipfel20, goodman2020substrate} and strain\,\cite{Uddin22}. In particular, EEA was demonstrated to be highly inefficient in hBN-encapsulated WS\textsubscript{2}, displaying an EEA rate of $0.004\ \text{cm}^2\ \text{s}^{-1}$ compared to $0.4\ \text{cm}^2\text{s}^{-1}$ for samples deposited on SiO\textsubscript{2} substrates\,\cite{Zipfel20}. The large EEA rate in TMD/SiO\textsubscript{2} structures has been suggested to partially originate from the relaxed energy- and momentum-conservation due to the strong dielectric disorder in these structures \cite{Zipfel20}. More recently, a microscopic model showed that the band structure in TMD/SiO\textsubscript{2} structures is more favourable for EEA due to the optimal energy of the higher-lying exciton state \cite{PRXGlazov, lin2021narrow, erkensten2021}.
In addition, the modification of the band structure via tensile strain has been recently shown to efficiently suppress EEA in monolayer WS\textsubscript{2} on a SiO\textsubscript{2} substrate\,\cite{Uddin22}.

\subsection{Trion and biexciton dynamics} 
In the presence of doping, trions govern the dynamics in TMDs. A lot of research  is currently focusing on characterizing and understanding the dynamics of trion formation and thermalization.
Initial studies resolved the formation of trions in pump-probe spectroscopy and determined a formation time of about 2 ps at cryogenic temperatures\,\cite{singh16,godde16}.
At the same time, the spin-valley depolarization dynamics was extensively studied\,\cite{mak13,Wang14,Plechinger16,singh16b} with key studies reporting a very long-lived valley coherence for trions in MoS\textsubscript{2}\,\cite{mak13} and WSe\textsubscript{2}\,\cite{singh16b}.
Recently, the role of excited trion states has been investigated. In particular, it was found that 2s excitons bound to a doping charge can relax into the ground 1s exciton state on a timescale of 60 fs, transferring the excess energy to the doping charge in a process denoted as autoionization\,\cite{wagner20}.

Trions exhibit a characteristic recombination mechanism in which one electron-hole pair recombines and the trion's center-of-mass momentum is transferred to the remaining charge carrier, which then becomes free (cf. Fig.\,\ref{fig:4}D). This phenomenon is known as the electron recoil effect\,\cite{esser01} and leads to an asymmetric broadening of the trion PL, as opposed to the symmetric broadening usually displayed by bright-exciton PL. Importantly, the broadening of the low-energy tail of the trion PL is determined by the temperature of the trion gas\,\cite{christopher17,zipfel22}. While the electron recoil effect has been known for a long time\,\cite{esser01,ross13,christopher17}, it has been exploited to extract the trion temperature and resolve the cooling of the trion population after an optical excitation only recently\,\cite{zipfel22}.
In Fig.\,\ref{fig:4}C time snapshots of the trion PL in MoSe\textsubscript{2} are shown\,\cite{zipfel22}. Immediately after the optical excitation, the trion PL displays a broad low-energy tail that shrinks on a 10 ps timescale. The shrinking is a direct manifestation of the trion temperature cooling down and reaching an equilibrium with the thermal bath of lattice phonons (cf. Fig.\,\ref{fig:4}E). These measurements can be systematically used to study how the trion cooling dynamics is affected by external knobs, such as lattice temperature and doping. In fact, the cooling time was found to be approximately constant ($\sim10\ \text{ps}$) up to 40 K. At higher temperatures, the thermal activation of optical and zone-edge acoustic phonons enhances the trion-phonon scattering, resulting in faster cooling rates. While trion-phonon scattering is the dominant cooling mechanism up to doping levels corresponding to a hole density of $3 \cdot 10^{11}\ \text{cm}^{-2}$, scattering between trions and free carriers becomes relevant at larger doping levels and further reduces the cooling time.
The microscopic mechanisms governing trion-phonon interaction have been recently studied in a theoretical work\,\cite{Perea22}.

Understanding the main trion recombination pathways is crucial for an efficient doping tunability of optical devices. The trion recombination time in MoSe\textsubscript{2} monolayers was found to change significantly with temperature, varying from 15 ps at 7 K to sub-picosecond timescales above 100 K\,\cite{robert16}. This strong temperature dependence was suggested to be caused by the thermal activation of phonon-assisted trion dissociation into an exciton and a free electron. Recent studies have estimated the non-radiative recombination time of trions in molybdenum-based TMDs to be typically around 50-100 ps, i.e. competing or even dominating over the radiative lifetimes of 100 ps\,\cite{fang19} and 100 ns\,\cite{lien19} that have been reported in MoSe\textsubscript{2} and MoS\textsubscript{2}, respectively.
The significant discrepancy between trion recombination times in different studies highlights the need for a better understanding of this process.
In particular, future studies should be carried out at carefully controlled experimental conditions and be accompanied by theoretical models accounting for both radiative and non-radiative recombination mechanisms.

At high exciton densities, more complex many-particle compounds, such as biexcitons, can become relevant. The spin-valley configuration of optically accessible biexciton states, as well as the characteristic quadratic dependence of the PL with pump power, has been reported\,\cite{you2015observation,barbone18,Ye18,steinhoff2018biexciton}.
In particular, bright biexcitons were found to be composed of one KK and one K'K' exciton with the opposite spin configuration in WSe\textsubscript{2} monolayers\,\cite{steinhoff2018biexciton}.
More recently, biexciton recombination and exciton-exciton scattering were predicted to be largely tunable by external magnetic fields\,\cite{katsch2020optical}. In the presence of an in-plane or tilted magnetic field, the spin selection rules for biexciton recombination and exciton-exciton interaction are relaxed, allowing for more recombination and scattering channels.

\begin{figure*}[t!]
\includegraphics[width=0.9\linewidth]{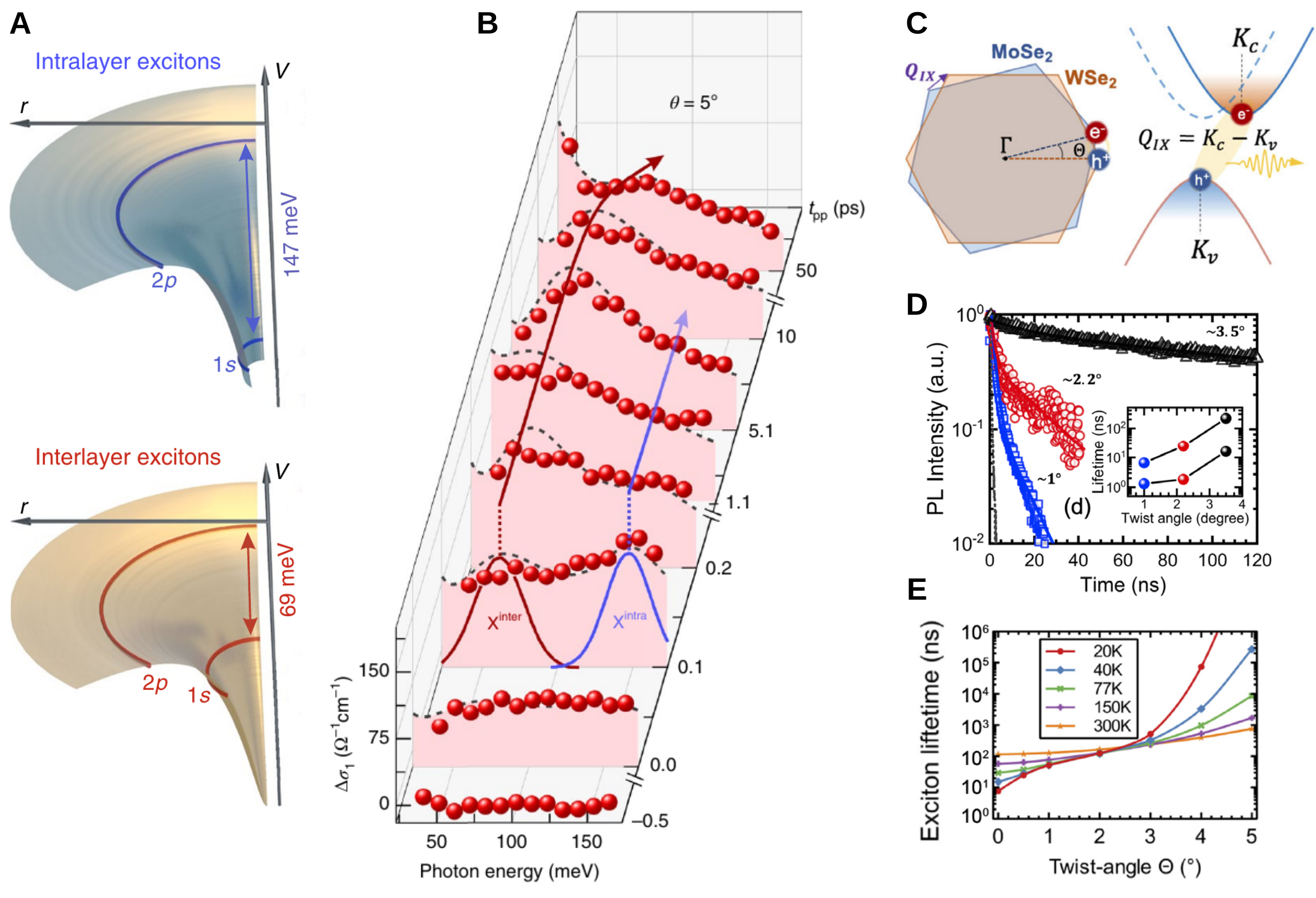}
\caption{Exciton dynamics in TMD heterostructures
\textbf{A}: Intra- and interlayer Coulomb potential together with the energetic position of the 1s and 2p exciton states in a tungsten-based vdW heterostructure.
\textbf{B}: Time- and energy-resolved pump-induced change of the optical conductivity in an almost aligned (5° twist angle) heterostructure illustrating the transition from an intra- to an interlayer exciton.
\textbf{C}: Schematic representation of the MoSe\textsubscript{2} and WSe\textsubscript{2} Brillouin zones twisted with respect to each other (left panel). The finite twist angle makes the interband transition in this heterostructure indirect (right panel).
\textbf{D}: Photoluminescence intensity in  MoSe\textsubscript{2}/WSe\textsubscript{2} heterostructures as a function of time for different twist angles. The time evolution of the PL is fitted with a biexponential function and the extracted lifetimes are displayed in the inset.
\textbf{E}: Predicted exciton lifetime as a function of twist angle for different temperatures.
Panels adapted with permission from: \textbf{A}-\textbf{B}, Ref.\,\cite{Merkl19}, Springer Nature Ltd.; \textbf{C}-\textbf{E}, Ref.\,\cite{choi21}, APS.
}
\label{fig:5}
\end{figure*}

\subsection{Exciton dynamics in TMD heterostructures}
TMD heterostructures offer a platform to spatially separate charge carriers. The resulting charge-separated state, i.e. the interlayer exciton or charge-transfer exciton, has characteristic properties that can be utilized to modify the optical response and transport properties of the material. In this section, we summarize the recent developments in the understanding of the charge transfer process in TMD bilayers. Moreover, we discuss the recombination dynamics in these systems and the impact of the twist angle between the two TMD layers.

Following the characterization of exciton formation and thermalization in monolayer TMDs, efforts have been directed towards investigating the dynamics of charge transfer and the formation of spatially separated interlayer excitons in vdW heterostructures\,\cite{jin18,Plankl21,schmitt21}.
The formation dynamics of interlayer excitons has been tracked in tungsten-based heterostructures using optical-pump THz-probe spectroscopy at room temperature\,\cite{Merkl19}. 
The density of intra- and interlayer excitons can be accessed by measuring the THz conductivity at the 1s-2p transition energy of each exciton species (cf. Fig.\,\ref{fig:5}A). The time-resolved THz conductivity is shown in Fig.\,\ref{fig:5}B. The initial optical excitation generates intralayer excitons in the WSe\textsubscript{2} layer, which becomes manifest as a peak in the THz spectrum at the 1s-2p transition energy of 150 meV \cite{Merkl19}. At the same time, a low-energy peak at approximately 70 meV originating from interlayer excitons emerges, indicating that the charge-transfer occurs on a sub-picosecond time scale.
After the optical excitation, the intralayer exciton peak vanishes, while the interlayer exciton resonances becomes more prominent as the whole exciton population relaxes into this energetically favourable state. These features reflect the transfer of electrons from the WSe\textsubscript{2} layer to the WS\textsubscript{2} layer, resulting in the formation of interlayer excitons with an excess energy that is later dissipated into the phonon bath\,\cite{Ovesen19}.
Intriguingly, the time scale of charge-transfer strongly depends on the stacking angle of the vdW heterostructure. This effect has been shown to arise from the momentum offset between intra- and interlayer exciton dispersions (cf. Fig.\,\ref{fig:5}C).

The charge transfer process in WSe\textsubscript{2}/MoS\textsubscript{2} moiré lattices has been recently resolved directly in tr-ARPES experiments\,\cite{schmitt21}, revealing an ultrafast sub-50 fs tunneling time scale. It was found that the charge transfer is a two-step process, where optically excited WSe\textsubscript{2} intralayer excitons first scatter with phonons into the strongly hybridized momentum-dark K{\textLambda} excitons (where electrons are delocalized over both layers). In a second step, these hybridized excitons scatter into the energetically lowest interlayer exciton states, where the electron is located in the MoSe\textsubscript{2} layer. The charge transfer process is sensitive to temperature as it is mediated by phonons, and strongly depends on the stacking as this determines the energy landscape of excitons \cite{schmitt21,hagel2021exciton,meneghini2022ultrafast}.

Apart from vdW heterostructures consisting exclusively of individual TMD monolayers, TMD-graphene structures have also been extensively studied due to their technological relevance\,\cite{massicotte16,he14b,yuan18,lorchat20,aeschlimann20,krause21,fu21}.
In particular, TMD and graphene can be stacked together to form optoelectronic devices, such as photodetectors or solar cells, where the TMD acts as an optical absorber and graphene acts as a transparent contact\,\cite{yu2013highly,britnell2013strong}.
In a recent work, the charge transfer in WS\textsubscript{2}/graphene structures after an optical excitation of excitons in the WS\textsubscript{2} layer was found to occur on a sub-picosecond time scale\,\cite{aeschlimann20}.
Such ultra-fast charge transfer from TMDs to graphene has recently been shown to lead to a complete neutralization of the TMD and thus to a filtering of the PL spectra displayed by these structures\,\cite{lorchat20}. Instead of the complex PL spectra of TMD monolayers, which involve trions and charged biexcitons, TMD-graphene heterostructures exhibit single narrow PL peaks arising from radiative recombination of neutral excitons.
Furthermore, the asymmetry between the electron and hole transfer leads to a charge-separated state that can last between 1 ps\,\cite{aeschlimann20,krause21} and 1 ns\,\cite{fu21}. The charge-transfer process in WS\textsubscript{2}/graphene was suggested to occur directly at the band intersection around the K point of the two layers. Crucially, the higher energy barrier and suppressed tunneling strength for electrons hinders their tunneling from WS\textsubscript{2} into the graphene layer, whereas holes can efficiently tunnel and relax into the graphene Dirac cone\,\cite{krause21}. In addition, defects in the TMD layer have been suggested to modify the charge separation lifetime by trapping the electrons into localized states\,\cite{krause21,fu21}.

While radiative recombination in TMD heterostructures has been studied for a few years\,\cite{Rivera15,Miller17,jiang18,nagler17,jauregui19,merkl2020twist}, only recently its twist-angle dependence has been characterized and understood on a microscopic footing\,\cite{choi21}. Importantly, the twist angle determines the momentum offset between the conduction band minimum and the valence band maximum of the two layers (cf. Fig.\,\ref{fig:5}C). As the transition becomes more indirect, radiative recombination becomes less efficient, resulting in a longer exciton lifetime for larger twist angles (cf. Fig.\,\ref{fig:5}D). Moreover, the moiré potential significantly impacts the radiative lifetimes by relaxing the momentum conservation of the recombination process and modifying the energy landscape of moiré excitons. The radiative lifetime was found to increase by one order of magnitude when increasing the twist angle from 1° to 3.5° in MoSe\textsubscript{2}/WSe\textsubscript{2} heterostructures\,\cite{choi21}. Due to the different energy landscape of moiré excitons at small and large twist angles, the radiative lifetime displays an opposite temperature dependence in these two regimes. At small twist angles, the light cone becomes depleted for increasing temperatures resulting in a longer lifetime, while at large twist angle the opposite occurs (cf. Fig.\,\ref{fig:5}E).

Non-radiative recombination via Auger scattering has been suggested to be suppressed in TMD homobilayers due to the hybrid nature of the exciton, which coexists in both layers\,\cite{yuan2015exciton,siday2022ultrafast,rivera2018interlayer,Merkl19}. In fact, the EEA rate was found to decrease from $0.4\ \text{cm}^2\text{s}^{-1}$ in WSe\textsubscript{2} monolayers to $0.006\ \text{cm}^2\text{s}^{-1}$ in bilayers\,\cite{yuan2015exciton}.
Intriguingly, Auger scattering has been suggested to be enhanced with an out-of-plane electric field that polarizes excitons and effectively increases the exciton-exciton interaction\,\cite{wang2018electrical}.
A more recent work has attributed the bimolecular recombination observed at high densities in WSe\textsubscript{2} bilayers to the recombination of unbound electron-hole pairs\,\cite{siday2022ultrafast}. This has been exploited to resolve the Mott transition, which was measured to occur around $7.4 \cdot 10^{12}\ \text{cm}^{-2}$ in such structures. Above this density, electron-hole pairs are unbound and the dynamics is governed by an electron-hole plasma\,\cite{Steinhoff17}.


\section{Exciton transport}
Exciton dynamics in atomically thin semiconductors has been intensely studied for more than a decade, following the first successful exfoliation of TMD monolayers. However, the spatiotemporal dynamics of excitons has been put in the focus of 2D materials research only recently\,\cite{Cadiz17, Kulig18,Perea19,sun2022excitonic, yuan2020twist}.
The understanding of exciton propagation is crucial for potential applications of TMD monolayers and heterostructures that could rely on the controlled transport of excitons in the material. Importantly, while excitons are charge-neutral particles, their spatial propagation can be directed by potential gradients generated by, e.g., inhomogeneous strain\,\cite{Rosati21e}.
Being able to track excitons in space and time is of key importance for the successful development of TMD-based devices as it enables the opportunity to directly manipulate exciton currents. In this section, we summarize the recent developments on exciton transport, with focus on exciton diffusion and funneling in TMD monolayers as well as in vertical and lateral heterostructures. 

\begin{figure*}[t!]
\includegraphics[width=\textwidth]{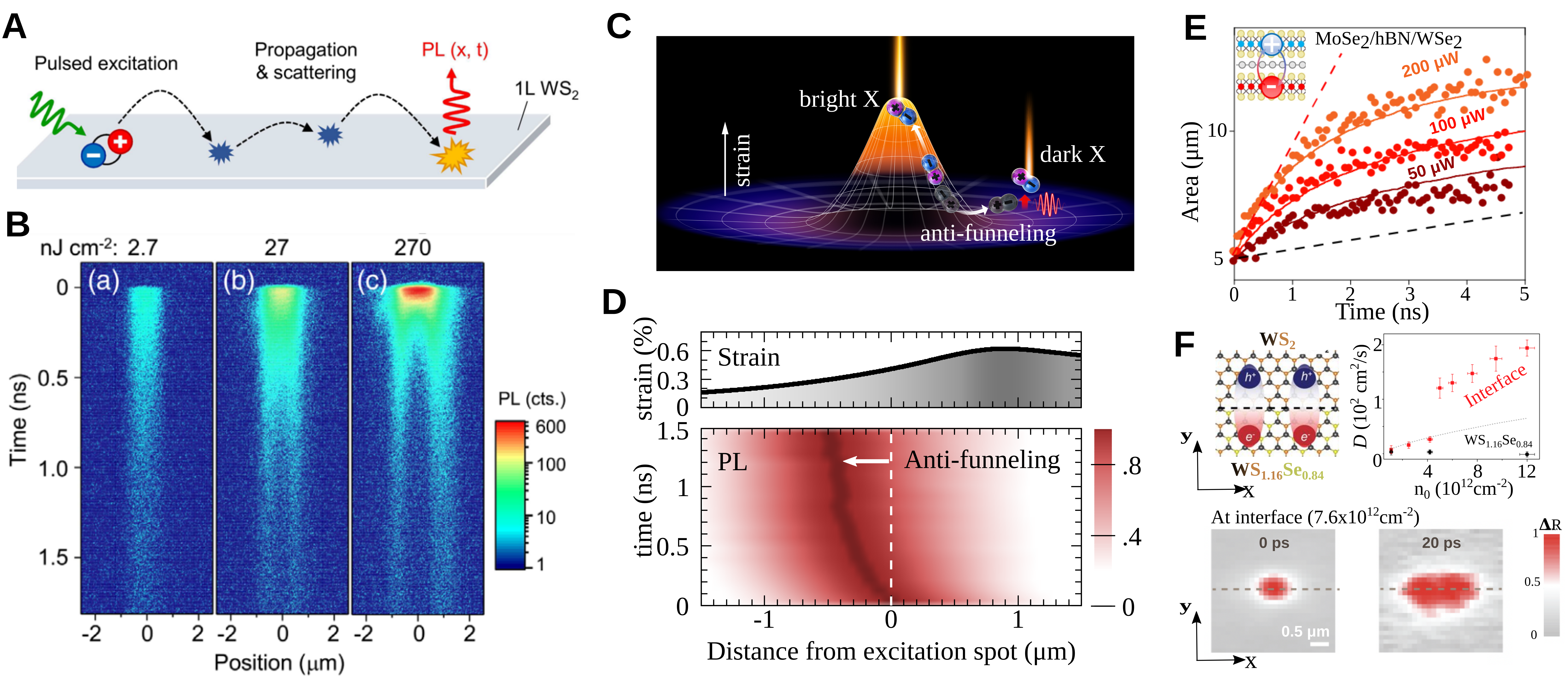}
\caption{Exciton transport in TMD monolayers and heterostructures.
\textbf{A}: Schematic illustration of exciton propagation and recombination in TMD monolayers.
\textbf{B}: Space- and time-resolved photoluminescence (PL) for a supported WS$_2$ monolayer revealing the formation of a halo-like profile at large pump power densities.
\textbf{C}: Schematic illustration of exciton funneling of bright excitons and anti-funneling of dark excitons.
\textbf{D}: Strain profile (top) and spatiotemporal evolution of the exciton PL (bottom) in monolayer WS$_2$, showing the anti-funneling of excitons towards regions of low strain attributed to the drift of dark K{\textLambda} excitons.
\textbf{E}: Time evolution of the exciton area (spatial variance) for different pump powers, demonstrating the anomalous diffusion of interlayer excitons in MoSe$_2$-hBN-WSe$_2$ heterotrilayers.
\textbf{F}: Schematic illustration of CT excitons in lateral heterostructures (top left). Differential reflection maps at 0 ps and 20 ps along the interface (bottom) demonstrating exciton diffusion in lateral heterostructures. Extracted diffusion coefficient as a function of electron-hole density (top right), revealing an abrupt increase of the spatial-spread speed after high-power excitation at the interface.
Panels adapted with permission from: \textbf{A}-\textbf{B}, Ref.\,\cite{Kulig18}, APS; \textbf{C}-\textbf{D}, Ref.\,\cite{Rosati21e}, AAAS under a Creative Commons license; \textbf{E}, Ref.\,\cite{sun2022excitonic}, Springer Nature Ltd.; \textbf{F}, Ref.\,\cite{Yuan21}, AAAS under a Creative Commons license.  
} 
\label{fig:6}
\end{figure*}

\subsection{Exciton diffusion}
Excitons are commonly tracked in both space and time by means of spatiotemporal photoluminescence measurements\,\cite{Cadiz17, Kulig18, Zipfel20, Uddin20}. As schematically shown in Fig.\,\ref{fig:6}A, excitons form after optical excitation, and then propagate and scatter. 
After some time, a fraction of excitons recombines radiatively, emitting light at a finite distance from the initial excitation spot.
In Fig.\,\ref{fig:6}B(a), a streak camera image of the PL for a low excitation density is shown, revealing a signal which decays with time and broadens in space.
Quantitatively, in the low-excitation regime, the PL intensity $I_{\mathrm{PL}}(x,t)$, being directly proportional to the exciton density $n(x,t)$, can be fitted to a Gaussian $\sim \mathrm{exp}(-x^2/\sigma^2_t)$, where $\sigma^2_t=4Dt+\sigma_0^2$ is the spatial variance and $D$ is the diffusion coefficient.
The spatial variance changes linearly with respect to time and the Gaussian shape of the exciton density $n(x,t)$ is retained for all times $t$. This spatiotemporal behavior can be explained by the Fick's law for  diffusion\,\cite{pathria2016statistical} $\partial_t n=D\nabla^2n$,  which can be recovered from the full state-dependent Boltzmann equation of the spatiotemporal exciton dynamics in the limit of fast scattering and quasi-thermalized local distributions\,\cite{Hess96,Rosati20}.
In this thermalized regime and in the limit of state-independent scattering times $\tau$, the diffusion coefficient becomes $D=k_BT \tau/M$ with $T$ and $M$ being temperature and total exciton mass, respectively\,\cite{Hess96,Rosati20}. More recently, low-temperature deviations from this simple temperature dependence of $D$ have been observed and attributed to quantum interference effects\,\cite{Glazov20,Wagner21}. Therefore, the transport behavior observed in Fig.\,\ref{fig:6}B(a) at low excitation powers is commonly referred to as conventional or regular diffusion. 
This takes place after the excitons have reached a thermal equilibrium, while, before that, excitons exhibit ballistic propagation ($\sigma^2_t\propto t^2$) and transient diffusion\,\cite{Rosati20,Rosati21c}.

When increasing the excitation density in the considered WS$_2$ monolayer, a halo-like pattern emerges in the spatial PL map which indicates that excitons quickly move out of the excited high-density region\,\cite{Kulig18}, cf. Fig.\,\ref{fig:6}B(b)-(c). The now time-dependent effective diffusion coefficient $D_{\mathrm{eff}}(t)=\frac{1}{4}\frac{d}{dt}\sigma^2_t$ was observed to increase from $0.3 \ \mathrm{cm}^2/\mathrm{s}$ at low excitation densities (below $n_x=10^{9}$ $\mathrm{cm}^{-2}$) up to $20 \ \mathrm{cm}^2/\mathrm{s}$ at a moderate exciton density of $n_x=10^{11}$ $\mathrm{cm}^{-2}$ (cf. Fig.\,\ref{fig:6}B(c)). 
The microscopic mechanism governing the formation of exciton halos and the unconventional non-linear diffusion observed in TMD monolayers at elevated exciton densities\,\cite{Kulig18, Cordovilla19, Perea19, park21} has been ascribed to the formation of strong spatial gradients in the exciton temperature\,\cite{Perea19,Glazov19}. Such temperature gradients are caused by the scattering of excitons with hot phonons that are created during the relaxation cascade of high-energy Auger-scattered excitons. These high-energy excitons are the result of efficient Auger scattering.
While the gradient in the exciton temperature explains the observed diffusion and halo formation at room temperature, another phenomenon---the phonon wind effect---has been predicted to play a relevant role in the formation of halos at cryogenic temperatures \,\cite{Glazov19}. In this effect, phonons created during the relaxation of Auger-scattered excitons propagate ballistically, dragging excitons out of the excitation spot. 
Similar non-linear diffusion\,\cite{yu20} and halo formation\,\cite{yu19} have been reported in monolayer MoS$_2$ and were suggested to arise from screening of the exciton-phonon interaction by trapped charges and from the formation of an electron-hole liquid, respectively.

Apart from excitation density, exciton diffusion can also be controlled via dielectric engineering\,\cite{Fu19,Raja19,Zipfel20,Li21}, strain\,\cite{Harats20, Rosati21a,Uddin22} or gating\,\cite{Uddin20, cheng2021observation}. The energy landscape of excitons, consisting of bright and dark exciton states, is highly sensitive to both strain\,\cite{Peng20,Niehues18} and changes in the dielectric environment\,\cite{Raja17}. Comparing TMD layers deposited on a SiO$_2$ substrate with hBN-encapsulated samples, the latter display weak dielectric disorder leading to a faster diffusion\,\cite{Zipfel20,Raja19, Li21}.
Moreover, strain can alter the relative positions of different excitonic valleys\,\cite{Khatibi18,Aslan18}, resulting in a strain-controllable closing of scattering channels, giving rise to diffusion coefficients increased
by a factor of 3 with about 0.6\% biaxial strain
\cite{Rosati21a}. Finally, gating can be used to disentangle neutral exciton diffusion from contributions to the diffusion due to unintentional doping and trions\,\cite{Uddin20, cheng2021observation}.  

In recent years, also the transport and propagation of quasiparticles beyond excitons has been studied extensively. Interestingly, the propagation of exciton-polaritons is strongly influenced by their excitonic or photonic nature. For example, exciton-polaritons in WSe\textsubscript{2} waveguides were observed to propagate at a velocity of $0.017c$, where $c$ is the speed of light\,\cite{mrejen2019transient}. In other words, photons are significantly slowed down when they interact strongly with excitons.
Moreover, thanks to the long propagation lengths of exciton-polaritons with a small effective mass, the optical valley Hall effect has been demonstrated in MoSe\textsubscript{2} embedded in a microcavity\,\cite{lundt19}.
Trions also provide an intriguing platform for charge transport, as they acquire the large oscillator strength of the exciton and the charge of the extra electron or hole. In fact, the drift of trions under the influence of an electric field has recently been observed\,\cite{cheng2021observation}. Here, a slow diffusion coefficient of $0.47\ \text{cm}^2/\text{s}$ and a drift velocity of $7400\ \text{cm}/\text{s}$ were reported in a WS\textsubscript{2} monolayer at cryogenic temperatures.
Other studies have investigated trion diffusion and reported substantially different diffusion coefficients ranging from the $0.47\ \text{cm}^2/\text{s}$ mentioned above up to $18\ \text{cm}^2/\text{s}$\,\cite{Uddin20,Kim21}, with signatures of even faster diffusion at cryogenic temperatures\,\cite{park21}.
While a recent theoretical work has studied the impact of trion-phonon scattering on trion transport\,\cite{Perea22}, a microscopic understanding of trion diffusion in different regimes is still lacking.

\subsection{Exciton funneling}
Current nanoelectronics relies on controlling the charge transport within the device. While for charge carriers this can be achieved by applying electric fields, other mechanisms are required to spatially guide neutral excitons. 
Strain engineering provides a way of manipulating exciton propagation in TMDs, since the band structure of these materials is remarkably sensitive to strain\,\cite{Aslan18,Khatibi18}. In particular, strain is seen to induce significant energy shifts of exciton resonances\,\cite{Niehues18,Peng20}.
Localized strain profiles as obtained e.g. via ripples\,\cite{Castellanos13}, bubbles\,\cite{Tyurnina19,Carmesin19,Darlington20}, pillars\,\cite{Branny17,Palacios17}, or patterned substrates\,\cite{Feng12,Kern16}, result in
space-dependent exciton energies $E\equiv E(\textbf{r})$ as reflected by spatially-resolved spectra.
Since excitons move toward the spatial positions where their energy is minimal,
their transport after thermalization can typically be described by a drift-diffusion equation
\begin{equation}
\partial_t n(\bm{r}, t)=\nabla\cdot ( D \nabla n)+\nabla( \mu n \nabla V(\bm{r}, t )) \ ,
    \label{dde}
\end{equation}
where $n(\textbf{r},t)$ is the spatially and time-dependent exciton density and $\mu$ is the exciton mobility, which can be approximated at low densities as $\mu\approx{\frac{D}{k_B T}}$. 
The first term in Eq. \eqref{dde} corresponds  to the Fick's law of diffusion\,\cite{pathria2016statistical}. 
The second term describes the drift induced by a potential $V$, which could have different microscopic origin, e.g. inhomogeneous strain,
gating or repulsive dipole-dipole interactions in heterostructures\,\cite{unuchek2019valley,sun2022excitonic, PhysRevMaterials.6.094006}.
In the case of locally-strained monolayers, the drift force is caused by  strain-induced variations of  exciton energy $\nabla E(\textbf{r})$, cf. Fig.\,\ref{fig:6}C.

The spectral analysis of time-integrated photoluminescence in rippled few-layer MoS$_2$ revealed how localized strain profiles act as excitonic traps\,\cite{Castellanos13}. 
In particular, the 
relation between photoluminescence resonance and strain reveals that excitons funnel before recombining\,\cite{Castellanos13}.
Excitons funnel towards regions with maximum strain, which results in an enhancement of the PL intensity in these regions\,\cite{Li15funnell,Branny17}. 
Recently, the exciton drift has been experimentally tracked by means of space- and time-resolved photoluminescence measurements\,\cite{Cordovilla18,Moon20,Rosati21e}.
Exciting far away from a strain profile induced e.g. by a pillar, the PL becomes broader in space without shifting its central position\,\cite{Cordovilla18}. In other words, only diffusion is observed (without drift/funneling) due to the absence of local strain profiles close to the excitation spot\,\cite{Cordovilla18}. The situation differs significantly when the excitation is performed in strained regions, i.e. close to the pillar. In this case, the time-resolved PL profiles reveal exciton motion toward the center of the pillar, where the strain is maximized, by hundreds of nanometers in the first nanosecond\,\cite{Cordovilla18}.
Space- and time-resolved PL has been applied also to strain profiles dynamically altered by a tip\,\cite{Moon20}. In this way it has been shown how the tip-induced strain can literally steer the motion toward the maximum-strain positions\,\cite{Moon20}.

More recently, exciton anti-funneling has been reported\,\cite{Rosati21e}, with excitons moving toward regions of low strain, see Fig.\,\ref{fig:6}D.
This surprising behaviour was observed in WS$_2$ deposited on a borosilicate substrate with an array of polymer micropillars, in contrast to analogous experiments performed on MoSe$_2$\,\cite{Rosati21e}. This can be understood in terms of the rich exciton landscape in TMD monolayers. While the energies of bright (KK) states decrease with increased strain, the opposite holds for momentum-dark K{\textLambda} intervalley excitons\,\cite{Aslan18,Khatibi18}. 
The strain-induced variation $\nabla E_v(\textbf{r})$ of the energy $E_v(\textbf{r})$ in the exciton valley $v=\mathrm{KK}, \mathrm{K\Lambda}$ thus induces opposite drift forces for bright (KK) and dark (K{\textLambda}) exciton species. 
This complex interplay between bright and dark exciton propagation can be described by generalizing the drift-diffusion equation \eqref{dde} to valley-dependent exciton densities $n_v(\textbf{r},t)$ and energies $E_v(\textbf{r})$, and including the thermalization between exciton populations in different valleys\,\cite{Rosati21e}. 
In this way it has been predicted that the propagation of K{\textLambda} excitons, which represent the majority of the exciton population in WS$_2$ monolayers, follows the experimentally-observed anti-funneling behaviour, cf.  Fig.\,\ref{fig:6}D. Furthermore, the initially faster propagation reflects the larger diffusion coefficients $D(x)$ in the corresponding spatial positions\,\cite{Rosati21e}, as the diffusion coefficients depend on strain, i.e. $D\equiv D(s(\textbf{r}))$, due to opening/closing of intervalley scattering channels\,\cite{Rosati21a}.
While excitons in WS\textsubscript{2} show an anti-funneling behaviour, regular exciton funneling has been observed in MoSe$_2$, where, in contrast to WS\textsubscript{2}, most of the exciton population lives in the energetically-lower bright KK states\,\cite{Rosati21e}.

Similarly, driven by the momentum-dark exciton population, excitons have been found to funnel into dielectric inhomogeneities in WSe\textsubscript{2} bilayers\,\cite{Su22}. 
Furthermore, spin-dark excitons have been recently reported to show the same funneling behaviour as KK excions, i.e. drifting toward regions with high strain\,\cite{Gelly22}.
Strain profiles and the resulting funneling can also be engineered by nanoscale tips\,\cite{Moon20,Gelly22,Koo21} and surface acoustic waves\,\cite{Datta21,Datta22}. 
In addition, local strain profiles can be used also to create quasi one-dimensional channels, e.g. by depositing the monolayer on top of 1D semiconductor nanowires\,\cite{Dirnberger21}. This generates a highly-anisotropic diffusion with large diffusion coefficients of $\approx 10 \ $cm$^2$/s in the direction of the channel at room temperature.
These studies exemplify how exciton transport in TMD-based devices can be engineered with strain\,\cite{Dirnberger21}.

\subsection{Exciton transport in TMD heterostructures}
After having discussed transport of excitons in TMD monolayers, we now turn to vertical and lateral heterostructures. Here, the already rich exciton landscape is further extended to spatially separated interlayer excitons, usually denoted as charge transfer (CT) excitons in lateral junctions.

\textbf{Vertical heterostructures:}
Here, interlayer excitons are composed of electrons and holes residing in different vertically stacked TMD layers and exhibit permanent out-of-plane dipole moments, cf. inset in Fig.\,\ref{fig:6}E. As a consequence, interlayer excitons display a strong dipole-dipole repulsion giving rise to a density-dependent renormalization of the exciton energy. Hence, dipole-dipole repulsion becomes manifest in PL spectra as a blue-shift of the exciton resonance at elevated densities\,\cite{nagler2017interlayer,yuan2020twist, unuchek2019valley}. 
As expected from Eq.\,\eqref{dde}, the density-dependent energy renormalization  gives rise to a drift force, resulting in a highly anomalous exciton diffusion, where the variance of the spatial distribution of excitons does not evolve linearly in time\,\cite{unuchek2019valley, yuan2020twist, sun2022excitonic, ivanov2002quantum, PhysRevMaterials.6.094006}.

In the case of MoSe$_2$-hBN-WSe$_2$ heterostructures, a super-linear dependence of the spatial variance $\sigma_t^2$ with respect to time has been recently observed at exciton densities $n_x>10^{12}$ $\mathrm{cm}^{-2}$\,\cite{sun2022excitonic}, cf. Fig.\,\ref{fig:6}E. At these high densities the potential gradient $\nabla V=\nabla(gn)$ is large and the drift-term in Eq. \eqref{dde} becomes dominant. Here, $g$ is governed by repulsive dipole-dipole interactions due to the large separation between electrons and holes forming  interlayer excitons in this particular heterostructure. Generally, the exciton-exciton interaction consists of two parts---one part stemming from direct dipolar exciton-exciton repulsion and the other part originating from the quantum-mechanical exchange interactions reflecting the fermionic character of excitons\,\cite{Schindler08, ciuti1998role,shahnazaryan2017exciton, erkenstenxx}. Importantly, when the separation between the two layers is similar or larger than the exciton Bohr radius, the exchange interaction---which can be negative and thus weaken the exciton-exciton interaction---is suppressed and the direct dipole-dipole repulsion dominates\,\cite{de2001exciton, kyriienko2012spin, baghdasaryan2022}. Moreover, the hBN spacer between the TMD layers does not only boost the dipole moment of the interlayer excitons, but it also suppresses the moiré potential which could trap excitons and affect their propagation in heterobilayers\,\cite{choi2020moire, yuan2020twist, wang2021diffusivity, huang2022excitons}.

A recent study on WS$_2$-WSe$_2$ bilayers shows that the diffusion of interlayer excitons is highly tunable with the twist angle, which modifies the moiré potential landscape\,\cite{huang2022excitons}.
In particular, the minima of the moiré potential act as traps, hindering the diffusion of excitons and effectively decreasing the diffusion coefficient. 
Moreover, it has been shown that moiré heterostructures are suitable candidates for realizing Bose-Hubbard models of interlayer excitons, where the twist-angle is used to control both the exciton-exciton interaction strength as well as the inter-site hopping amplitudes\,\cite{PhysRevB.105.165419, PhysRevB.106.035406}. These findings indicate that twisted TMD heterostructures could potentially be used as possible quantum simulators for bosonic many-body systems. 

\textbf{Lateral heterostructures:}
Recently, laterally stacked TMD monolayers have emerged as a promising new platform to study one-dimensional excitons. In these structures two different monolayers are grown beside each other and covalently stitched together in the plane\,\cite{Duan14,Gong14,Huang14,Li15,Heo15,zhang2017robust,Sahoo18}, cf. Fig.\,\ref{fig:6}F.
At the interface, the formation of spatially-separated CT excitons has been predicted\,\cite{Lau18,Yuan21}. 
While for interlayer excitons in vertical heterostructures the spatial separation of electron and hole is restricted by the layer distance,  separations of several nanometers are predicted for CT excitons thanks to the large device length\,\cite{Lau18,Yuan21}.
While exciting far away from the interface leads to the regular monolayer diffusion (black points in top-right Fig.\,\ref{fig:6}F), exciting close to the quasi one-dimensional channel formed at the interface results in intriguing transport properties\,\cite{Bellus18,Yuan21,Beret22,Shimasaki22}.
The different excitonic energies in the two monolayers drive the propagation of excitons towards and across the interface\,\cite{Bellus18,Yuan21,Shimasaki22}. Varying the central position of the excitation spots close to the interface of an hBN-encapsulated MoSe$_2$-WSe$_2$ lateral heterostucture, a non-trivial and non-monotonic behaviour of the diffusion length has been observed stemming from  different excitonic lifetimes and radiative recombination times of the two monolayers\,\cite{Beret22}.

A peculiar spatial expansion along the interface has been observed in a WSe$_2$-WS$_{1.16}$Se$_{0.84}$ lateral heterostructure\,\cite{Yuan21}. 
Here, an abrupt enhancement of the diffusion along the 1D channel is measured for excited densities above 5$\cdot 10^{12}\ \text{cm}^{-2}$, resulting in effective diffusion coefficients of the order of 100 cm$^2$/s, cf. the red points in  Fig.\,\ref{fig:6}F. This reflects a first-order Mott transition with the formation of a dense electron-hole plasma that diffuses quickly, leading to a broad electron-hole distribution along the interface already after few tens of picoseconds\,\cite{Yuan21}, cf. bottom row in Fig.\,\ref{fig:6}F. 
For intermediate excitation densities (from 1 to 4$\cdot 10^{12}\ \text{cm}^{-2}$) the effective diffusion coefficient increases with the excitation power up to values of few 10 cm$^2$/s, i.e. up to two orders of magnitude larger than the monolayer values obtained when exciting far away from the interface (cf. Fig.\,\ref{fig:6}F).
This increase in the effective diffusion coefficient has been attributed to the dipole-dipole repulsion between CT excitons\,\cite{Yuan21}, favored by the large spatial separations of several nanometers\,\cite{Lau18}.
These results put forward lateral heterojunctions as promising one-dimensional ``highways'' for excitons and unbound charge-carriers in an electron-hole plasma.


\section{Outlook}
The huge number of studies on mono- and multi-layered TMDs---highlighting the crucial importance of the rich and versatile exciton landscape for optoelectronic applications and fundamental research---has opened up many avenues that still remain unexplored in this field. Here, we outline some of these prospective research directions.

\textbf{Optics:}
While optical signatures of the exciton landscape including bright and dark excitons in both TMD monolayers and heterostructures have been extensively studied, the impact of different valley and spin configurations of more complex many-body compounds is still not completely understood.
In particular, previous studies have not addressed the potential relevance of trion states with charge carriers located around symmetry points other than the K or K' valleys. 
Although it is well known that excitons formed by electrons (holes) at the {\textLambda} ({\textGamma}) point are the energetically lowest or very close to the lowest-lying states in some monolayers and heterostructures\,\cite{Aslan18,Deilmann19,hagel2021exciton}, the energetic position of e.g. trion or biexciton states composed of charges in these valleys has remained in the dark.
The optical response of these higher-order charge complexes  could be further controlled by tuning the relative energetic position of the involved valleys with strain\,\cite{Aslan18,Khatibi18}.

The impact of strongly correlated states on the optical response of TMDs is a hot topic of research that still needs to be further explored. Very recently, signatures of many-body states beyond trions have been identified in reflectance spectra of heavily electron-doped WSe\textsubscript{2} monolayers\,\cite{li2021many}. While these signatures have been proposed to arise from 6- and 8-body exciton states interacting with the Fermi sea of doping charges\,\cite{van2022hexcitons}, it remains unclear why these many-body compounds are not experimentally observed in hole-doped samples.
The optical fingerprint of excitons interacting with a Wigner crystal of electrons has also been reported\,\cite{smolenski2021signatures}, providing a tool to investigate strongly correlated electron systems. A recent theoretical study has suggested to use terahertz light to directly probe the internal quantum transitions of the Wigner crystal itself \cite{brem2022terahertz}.
The experimental realization of correlated exciton phases such as exciton Wigner supersolids\,\cite{joglekar2006wigner,zhang20214} could shed light on the interplay between exciton-exciton interactions and their bosonic quantum statistics.

Regarding TMD polaritonics, one current focus in this field concerns utilizing polaritons as both a probe of, and as a potential means to control, microscopic processes in TMDs. For example, recent theoretical studies have explored how the shape of the polariton dispersion and modification of the effective phonon scattering matrix can drastically alter phonon scattering rates in comparison to bare excitons\,\cite{lengers2021phonon,ferreira2022microscopic}. Even more sophisticated microscopic theories will be necessary to fully describe newly reported hybrid exciton-photon-phonon states\,\cite{li2022hybridized}. Moving beyond monolayer exciton-polaritons provides a promising route towards quantum nonlinear devices. In addition to trion polaritons\,\cite{emmanuele2020highly}, interlayer exciton-polaritons represent an exciting combination of quantum tunnelling with strong light-matter coupling that possess a permanent dipole moment\,\cite{datta2021highly,louca2022nonlinear}. These dipolaritons offer enhanced nonlinearity as well as highly controllable polariton transport and condensation, making them relevant for potential quantum optoelectronic applications\,\cite{cristofolini2012coupling}.

\textbf{Dynamics:}
The recent observation of the ultrafast formation of momentum-dark excitons in tr-ARPES measurements\,\cite{Madeo20} represents a major step forward in the characterization and understanding of the exciton dynamics in TMD monolayers and heterostructures. Nevertheless, these recent experiments excited the TMD with linearly-polarized light and therefore simultaneously generated KK and K'K' excitons with opposite spin configurations. Therefore, the spin-relaxation via e.g. scattering with chiral phonons or intervalley exchange could not be distinguished from spin-conserving intervalley thermalization. The direct visualization and understanding of the main spin relaxation processes would be crucial for an accurate characterization of the limits and advantages of TMD-based spintronic devices.

Moreover, the first tr-ARPES studies on TMD heterostructures have been performed revealing crucial insights on the main charge transfer channels\,\cite{schmitt21} as well as indications of exciton localization resulting from superlattice-periodic moir\'e potential\,\cite{karni2022structure}. However, further studies of this kind involving different hetero-/homobilayers and varying crucial system parameters like twist-angle and temperature are necessary to confirm these first observations and to provide conclusive understanding of charge-transfer and localization dynamics. In particular, resolving the small moir\'e induced splittings of the ARPES signal in momentum and energy and preparing comparable samples with different stacking configurations provides a major experimental challenge. Furthermore, a microscopic model predicting the ARPES signatures of strongly correlated electronic states as well as superlattice umklapp-processes is still missing.

Exciton-exciton annihilation in TMD monolayers has been well characterized so far---the main scattering channels involved have been identified\,\cite{erkensten2021} and the tunability with external knobs has been explored\,\cite{goodman2020substrate,Zipfel20,erkensten2021,Uddin22}.
However, the understanding of Auger recombination of excitons in TMD heterostructures is still lacking. While it has been observed that EEA is severely weakened in homobilayers\,\cite{yuan2015exciton}, the microscopic origin of this suppression is unclear. More experimental and theoretical studies are needed to understand the microscopic mechanisms behind EEA and the tunability of this process in TMD heterostructures.

Studies on excitonic many-body complexes have so far focused on the optical fingerprint of these quasi-particles in absorption and PL spectra. Only recently the trion thermalization dynamics have been investigated by exploiting the temporal evolution of the PL line shape\,\cite{zipfel22}. Similar methods could be used to study the thermalization dynamics of other many-body complexes like polaritons and biexcitons.
Such experimental studies, together with appropriate theoretical models, would provide a better understanding of the main scattering mechanisms involved in the thermalization of photoexcited charges. The latter is crucial for an optimal design of efficient optoelectronic devices.

\textbf{Transport:}
In the recent years, the intrinsic mechanisms governing exciton diffusion and funneling have been well established. In particular, exciton diffusion is known to be controlled by exciton-phonon scattering at low exciton densities\,\cite{Perea19} and to be hindered by dielectric inhomogeneities and traps\,\cite{Raja19}.
It has also been understood that KK and {K\textLambda} excitons funnel in opposite directions of the strain gradient, although the reason for the unexpectedly large optical activation of {K\textLambda} excitons at room temperature has remained unclear\,\cite{Rosati21e}.
Furthermore, the understanding of the mechanisms governing the transport of more complex many-particle compounds is still lacking. The few reported diffusion coefficients for e.g. trions differ substantially in current literature\,\cite{Uddin20,Kim21,cheng2021observation,park21}, indicating the difficulty of accurately accessing the transport properties of many-particle states. Future studies should aim to thoroughly characterize diffusion of trions and other many-particle compounds consistently across different samples, temperatures, doping levels, and excitation densities in order to gain a better understanding of the underlying microscopic mechanisms.
We anticipate that, not only the main scattering mechanisms hindering propagation should be identified, but also the interplay between different many-particle compounds that might coexist (e.g. trions, excitons and single electrons) should be better understood. 
While research on excitons and higher-order charge complexes in atomically-thin materials has significantly advanced in the last decade, there is still a lot of new and exciting physics to be explored in the coming years.\\[0.5cm]

\noindent\textbf{Acknowledgements}\\
We acknowledge funding from the Deutsche Forschungsgemeinschaft
(DFG) via SFB 1083, and the European Union’s Horizon 2020 research and innovation program under grant agreement no. 881603 (Graphene Flagship).
R.P.C acknowledges funding from the Excellence Initiative Nano (Chalmers University of Technology) under its Excellence PhD program.

\end{document}